\documentclass[aps,pra,pifont,citesort,twocolumn]{revtex4}
\usepackage[utf8]{inputenc}
\usepackage{lineno}
\usepackage{color,soul,bm}
\usepackage{xcolor}

\usepackage{hyperref}
\hypersetup{
    colorlinks,
    linkcolor={violet},
    citecolor={teal},
    urlcolor={teal}
}

\usepackage{setspace}
\usepackage[utf8]{inputenc}
\usepackage{graphicx,amsmath,amssymb}
\usepackage{dcolumn}
\usepackage{bm}
\usepackage{times}
\usepackage{autobreak}

\newcommand{\bk}[1]{\left ( #1\right )}

\newcommand{\eqn}[1]{\begin{eqnarray} \newline #1 \end{eqnarray}}
\newcommand{\ee}{&=&}

\newcommand{\hs}{\hspace{0.2cm}}

\newcommand{\bra}[1]{\left \langle#1 \right |}
\newcommand{\ket}[1]{\left |#1\right \rangle}

\newcommand{\EV}[1]{\left < #1 \right >}

\newcommand{\nn}{\nonumber}
\newcommand{\half}{\frac{1}{2}}
\newcommand{\hmin}{H_{\mathrm{min}}}
\newcommand{\hmax}{H_{\mathrm{max}}}

\definecolor{darkgreen}{rgb}{0.0, 0.42, 0.24}

\definecolor{jens}{rgb}{0.1,0.5,0.1}
\definecolor{martin}{rgb}{0,0,1.0}

\usepackage{epstopdf} 
\usepackage{epsfig}
\newtheorem{Definition}{Definition}    
\newtheorem{Theorem}{Theorem}
\newtheorem{Protocol}{Protocol}
\newtheorem{Lemma}{Lemma}

\begin{document}

\title{Advantage of multi-partite entanglement for quantum cryptography\\ over long and short ranged networks}

\author{Janka Memmen${}^{1,2}$, Jens Eisert${}^{1,3,4}$, Nathan Walk${}^{1}$}
\affiliation{${}^{1}$Dahlem Center for Complex Quantum Systems, Freie Universit{\"a}t Berlin, 14195 Berlin, Germany \\
${}^{2}$Electrical Engineering and Computer Science Department,
Technische Universit{\"a}t Berlin, 10587 Berlin, Germany \\
${}^{3}$Helmholtz-Zentrum Berlin f{\"u}r Materialien und Energie, 14109 Berlin, Germany\\
${}^{4}$Fraunhofer Heinrich Hertz Institute, 10587 Berlin, Germany}

\date{May 29, 2025}

\begin{abstract}
The increasing sophistication of available quantum networks has seen a corresponding growth in the pursuit of multi-partite cryptographic protocols. Whilst the use of multi-partite entanglement is known to offer an advantage in certain abstractly motivated contexts, the quest to find practical advantage scenarios is ongoing and substantial difficulties in generalising some bi-partite security proofs still remain. We present rigorous results that address both these challenges at the same time. First, we prove the security of a variant of the GHZ state based secret sharing protocol against general attacks, including participant attacks which break the security of the original GHZ state scheme. We then identify parameters for a performance advantage over realistic bottleneck networks. We show that whilst channel losses limit the advantage region to short distances over direct transmission networks, the addition of quantum repeaters unlocks the performance advantage of multi-partite entanglement over point-to-point approaches for long distance quantum cryptography. 
\end{abstract}


\maketitle
\section{Introduction}
Multi-partite entanglement is a key ingredient in proposals for large-scale quantum communication networks - a quantum internet  \cite{Kimble:2008p5456,Wehner:2018cu,RodBook,WehnerVidick}. Substantial advances in network coding and the distribution of multi-partite entanglement in both theory    \cite{Kimble:2008p5456,Epping:2016hc,Epping:2017dx,Dahlberg:2018ku,Hahn:2018wq,Wallnofer:2019jw,Pirandola:2019bv,Khatri:2021kt,Das2021,Avis:2023jy} and experiment  \cite{Wang:2017ta,vanDam:2019fp,Proietti:2020ty,Thalacker:2021ic,Hermans:2022cd,Pickston:2023jw,Webb:2023gf,Roadmap} have brought these visions ever closer to reality. In particular, multi-partite entanglement can be used for quantum cryptographic tasks \cite{RevModPhys.74.145,NewQuantumCryptoReview} 
such as \emph{conference key agreement} (CKA), the distribution of a secret key to multiple trusted participants  \cite{Epping:2017dx,Grasselli:2018gk,Murta:2020de,Ottaviani:2019kp} and \emph{quantum secret sharing} (QSS), 
the distribution of shares of a secret key to several participants, an unknown subset of which may be malicious. In a so-called QSS $(N,k)$-threshold scheme, the dealer (Alice) has her key kept secret from any unauthorised set of $k-1$ participants (Bobs), whereas authorised sets of $k$ participants are able to resolve it  \cite{Shamir:1979cr}. The canonical multi-partite entangled state capable of accomplishing this task is the GHZ state, as first proposed by Hillery, Berthiaume and Bu{\v z}ek (HBB)  \cite{Hillery:1999tb} to implement an $(N,N)$ scheme. 

Cryptographic protocols necessitate both correctness and security. Proving the latter for QSS protocols based on GHZ states has turned out to be an extremely challenging task. In contrast to CKA, where all parties are trusted, the security of QSS protocols is most threatened by members of the protocol themselves, as they hold insider information on the protocol itself and actively take part in it. This is known as the participant attack and has first been pointed out in Ref.~\cite{Karlsson:1999ez}. As the participant attack involves the malicious parties making their announcements last, Ref.~\cite{Karlsson:1999ez} has specified the order of the announcement of measurement bases in the parameter estimation phase, but only for the case of $N=3$. Despite a substantial amount of follow up work \cite{Scarani:2001kd,Qin:2007fv,Xiao:2004fw,Chen:2007ul,Tittel:2001gn,Lance:2004do,Chen:2005fs,Schmid:2005hz,Gaertner:2007tt,Bogdanski:2009jo,Bell:2014ez,Grice:2015jw,Fu:2015df,Armstrong:2015he,Cai:2017cp,Zhou:2018hg,DeOliveira:2019ta,Markham:2008ta,Keet:2010cf,Marin:2013dc,Lau:2013gm} a more rigorous solution remained elusive. 
%
Eventually, two inequivalent proof methods were proposed in the asymptotic regime. Ref.~\cite{Williams:2019kb} proposed thwarting participant attacks for an arbitrary number of participants via a randomization of announcements of bases and outcomes, where every potentially untrusted subset has to go first at some point. This necessitates that a greater fraction of data be sacrificed for parameter estimation and implies a poor scaling of GHZ based QSS in the finite-size case \cite{ourqss}. In Ref.~\cite{Williams:2019kb}, it has also been essential that bases be chosen symmetrically to prevent players knowing a-priori what a given round will be used for, meaning that a QSS scheme could not be made arbitrarily efficient even asymptotically. An alternative asymptotic security proof based on the entropic uncertainty relation for a class of \emph{CV graph states} has been proposed \cite{Kogias:2017jz} recently generalised to the \emph{composable finite size framework} \cite{ourqss}. However, when applied to the original HBB protocol switching between the $X$ and $Y$ bases, this proof yields negative rates. 
In this work, we utilise a conceptually and practically simple modification to the original HBB protocol that nevertheless has a substantial impact:  the results of  Ref.~\cite{ourqss} can be non-trivially applied to the protocol which renders it invulnerable to the participant attack without requiring any additional classical data sacrifice to finally obtain a general, efficient, finite-size analysis for GHZ based secret sharing. A similar modification is also used in \cite{Li:2023ge} for to analyse multiplexed MDI-scheme that post-selects GHZ like correlations at a central relay station.

Both CKA and QSS can also be implemented by combining bi-partite QKD links, whose security is well established. Moreover, general point-to-point QKD can be made arbitrarily efficient by asymmetrical bases choice. This, along with the fact that the distribution of multi-partite entangled states is experimentally more demanding, raises doubts as to whether there is any advantage in pursuing GHZ-based protocols. However, in networks with bottlenecks multi-partite entanglement uses the network topology more efficiently and can generate an $N$-fold performance advantage of GHZ-based protocols for CKA compared to bi-partite QKD as shown in Ref.~\cite{Epping:2017dx}. Whilst the advantage persisted for a certain amount of depolarization on the channels and preparation noise, the biggest obstacle to quantum cryptography, transmission loss on the channels, has been neglected in their analysis. In a comparison of GHZ based protocols and bi-partite QKD, this is particularly important in that transmission loss affects GHZ based protocols significantly more due to the need for simultaneous successful transmission. 

To our knowledge, the only analyses including transmission loss are 
Ref.~\cite{ourqss} for QSS with CV graph states and 
Ref.~\cite{Grasselli:2022ci} for anonymous CKA. Crucially, once realistic losses are taken into account, the multi-partite entanglement advantage no longer grows as a function of player number. Instead it typically decreases with player number and vanishes altogether even for relatively small networks. We note that very recently a new protocol has been proposed that eschews entanglement and instead generalises twin-field QKD \cite{Lucamarini:2018if} with weak coherent pulses and an untrusted measurement relay to CKA and also achieves a speedup relative to a bi-partite implementation \cite{Carrara:2023da}. However, the advantage reported here also decreases montonically with the number of players. A recent experiment \cite{Pickston:2023jw} 
has also demonstrated a multi-partite advantage of CKA (a similar demonstration for anonymous CKA has also been carried out in
Ref.\ \cite{Webb:2023gf}). However, in this protocol a large, multi-partite state is always distributed and then subsequently transformed into either a bi-partite or multi-partite protocol on a smaller subset, which is not the most efficient way to distribute bi-partite entanglement. The two schemes are compared via secret bits per distributed entangled resource which naturally heavily penalises the bi-partite protocol, despite the fact that creating bi-partite entanglement is experimentally less challenging. A multi-partite advantage has  also been reported in 
Ref.\ \cite{Li:2023ge}, however, this work compares a multi-partite scheme that multiplexes a number of channels on each network link that increases inversely with the transmission probability (hence increasing exponentially with distance through fibre-optic networks) with a bipartite protocol that has only one channel per link.

In this work, we show an unambiguous
\emph{genuine advantage} for \emph{multi-partite entanglement} \cite{MultipartiteReview} 
using the \emph{standard benchmark of generated secret bits per network use}. Whilst the strategy of constantly generating a large resource state could have advantages from the perspective of flexibility in a future quantum internet, in the near term bits per network use will remain the most relevant figure of merit for assessing multi-partite advantages. 
We explicitly analyse both asymptotic and finite size rates for GHZ based QSS and compare them to ordinary point-to-point QKD in bottleneck networks including transmission loss as the primary decoherence problem as well as depolarising noise on channels. We identify parameter regimes in which the GHZ based protocol yields higher rates. Unsurprisingly, they ``melt'' and go away 
in the high loss regime and, for a fixed amount of loss, rapidly decrease and vanish as the number of network participants increases.  However, by adding quantum memories we unlock a genuine performance advantage for GHZ-based protocols for much further distances than before, which is robust for realistic finite block sizes and includes appropriate modelling of memory dephasing with parameters that are within the range of present day experiments. We find regimes of network and memory parameters such that, for a sufficiently high-quality memory, the multi-partite entanglement increases linearly with the player number for much larger networks ($N\approx 10$).

To put the significance of this work into a broader context, on the one hand, we solve the participant attack loophole and show an in-principle advantage of the use of GHZ entanglement in ($N,N$)-quantum secret sharing schemes, demonstrating a new functionality for these states. On the other hand, we make this advantage (and the analogous advantage for conference key agreement) plausible in a number of realistic settings, embedding the discussion in a framework respecting physically plausible desiderata that arise in settings of practical relevance. Specifically, by considering bottleneck networks with and without quantum memories we show that a multi-partite advantage is only possible for short transmission distances and small player numbers. However, the addition of memories `unlocks' the multi-partite advantage for substantially larger networks, in terms of both distance and number of players. We believe that these results, and future generalisations, can be used as meaningful benchmarks that simultaneously consider the quality of available resources for multi-partite entanglement generation and storage.


\section{Secret sharing rates} \label{secret_sharing_rates}
In order to correctly derive key rates, we need to group the participants into sets. The set of all Bobs is denoted as $\mathcal{B} = \{B_1,B_2,\ldots,B_n\}$. The set of all authorised or trusted subsets of $k$ Bobs is denoted as 
\begin{equation}
\mathcal{T} := \{T_1,T_2,\ldots,T_{\binom{n}{k}}\}
\end{equation}
for which $T_1 := \{B_1,B_2,\ldots,B_k\}$ and so on. Analogously, 
we define the set of all unauthorised or untrusted subsets of $(k-1)$ players \begin{equation}
\mathcal{U} := \{U_1,U_2,\ldots,U_{\binom{n}{k-1}} \}
\end{equation}
where, e.g., $U_1 := \{B_1,B_2,\ldots,B_{k-1}\}$ and so on. The \emph{extractable key} 
must essentially maximise over the possible dishonest parties information and minimise over possible honest parties information.


The asymptotic secret sharing rates emerge from the finite-size composable secure fraction in the limit of infinitely many rounds, perfect detection and information reconciliation \cite{ourqss}. In that limit so-called collective attacks, i.e., when the hostile parties act i.i.d.~(independently and identically distributed), are best to gain information without authorisation \cite{Renner:2007p1358,Renner:2009p1}. Defining the conditional 
von-Neumann entropy of $X_A$ given the quantum system $E,U_j$ (Eve plus the $jth$ unauthorised subset as
\eqn{S(X_A|E,U_j) = H(X_A)+\sum_{x_A}p(x_A) S(\rho_{E,U_j}^{x_A}) - S(E,U_j)}
with $H(X) := -\sum_{x}p(x)\log_2p(x)$ and $S(\rho) := -\mathrm{tr}\bk{\rho\log_2\rho}$
being the Shannon and von-Neumann entropies, respectively, we obtain the expected asymptotic formulas  \cite{Kogias:2017jz}
\eqn{K_{\mathrm{SS}} &:=& Y_{\mathrm{SS}}(\min_{U_j\in \mathcal{U}} S(X_A|E,U_j) - \max_{T_i\in\mathcal{T}} H(X_A|X_{T_i})).}
Note that here $X_{T_i}$ denotes whichever combination of measurements the trusted subset $T_i$ should make in order to make the best guess of Alice's variable $X_A$.

It is also interesting to compare this to the conference key agreement rate  \cite{Epping:2017dx} which, in our notation, is
\eqn{K_{\mathrm{CKA}}:= Y_{\mathrm{CKA}}(S(Z_A|E) - \max_{B_i\in \mathcal{B}}H(Z_A|Z_{B_i})) \label{Kcka}.}
The key points of difference are that, in a CKA protocol, there are never any players collaborating with Eve and also correlations between Alice and each Bob must be considered individually (since we do not want to have the Bobs needing to collaborate with each other) so the corresponding entropy is simply maximised over the set $\mathcal{B}$. Consequently, for GHZ based CKA, it is essential that the key is encoded in the Pauli $Z$ basis as this is the only basis the desired functionality can be achieved, i.e., where each player can hope to reconstruct Alice's measurement outcome by themselves. 

We will use an entropic uncertainty relation to bound the entropy of the dishonest parties. First, assume that Alice randomly switches between two non-commuting measurements $\mathbb{X}_A$ and $\mathbb{Z}_A$. Then, for each untrusted $(k-1)$-subset $U_j\in\mathcal{U}$, we can define the unique complementary subset $C_j \in \mathcal{C}$ of the $n-k+1$ remaining players where $C_j := \mathcal{B}\setminus U_j$. Now, by definition, the joint state of Alice, Eve and the sets $U_j$ and $C_j$ for all $j$ is pure. This means we can write down the entropic uncertainty relation  \cite{Berta:2010p1971,Furrer:2014ig},
\eqn{S\left(X_{A} | E,U_j\right)+S\left(Z_{A} | C_j\right) &\geqslant& c(\mathbb{X}_A, \mathbb{Z}_A).\label{eur}}
In the case of the measurements being Pauli 
measurements, we have that
\eqn{S(X_A|E,U_j) &\geqslant& 1 - h_2(Q_{Z_A|C_j})}
where $Q_{Z_A|C_j}$ is the bit error rate we have used that $H(Z_A|C_j)\geqslant S(Z_A|C_j)$ and that for binary outcome measurements $H(X) = h_2(Q_X):=-Q_X\log_2(Q_X) - (1-Q_X)\log_2(1-Q_X)$. Equivalently, 
we also have
\eqn{S(Z_A|E,U_j) \geqslant 1 - h_2(Q_{X_A|C_j}).}
Applying this to the CKA rate in (\ref{Kcka}) yields
\eqn{K_{\mathrm{CKA}} = Y_{\mathrm{CKA}}(1 - h_2(Q_{X_A|X_\mathcal{B}}) - \max_{B_i\in\mathcal{B}} h_2(Q_{Z_A|Z_{B_i}}))}
as previously shown (see, 
e.g., Eq.~(1) of Ref.~\cite{Proietti:2020ty}).
Turning to secret sharing, first note that for GHZ states we can only hope to achieve $(N,N)$-threshold schemes, since anything less than all players cooperating cannot possibly hope to reconstruct Alice's measurement. This is because GHZ states have the property that deleting a single subsystem leaves the remaining state maximally mixed. For the case $k=N$, there is only one trusted subset so we have $\mathcal{T} = \mathcal{B}$. Also, since each untrusted subset has $n-1$ players by definition the complementary subset is just one of the Bobs, i.e., $C_j = B_j$. In the original HBB protocol  \cite{Hillery:1999tb} Alice switches between Pauli $X$ and $Y$ measurements, so the key rate would be 
\eqn{K_{\mathrm{SS}}^{\mathrm{HBB}} = Y_{\mathrm{SS}}(1 - h_2(Q_{X_A|X_\mathcal{B}}) - \max_{B_j\in\mathcal{B}} h_2(Q_{Y_A|Y_{B_j}}))\label{hbbrate}.}
However, because of the nature of a GHZ state we know, even with no external decoherence, that Alice's Pauli $Y$ measurements look completely random to any single Bob so that $h_2(Q_{Y_A|Y_{B_j}})=1$ leading to a secret sharing rate in Eq.~(\ref{hbbrate}) 
that is always negative.

The solution is to keep encoding the key in the Pauli $X$ basis but make the check measurements in the Pauli $Z$ basis (see also \cite{Li:2023ge}). This makes the scheme essentially dual to the CKA protocol (where the key is in $Z$ and the check is in $X$). With this strategy our secret sharing rate would be
\eqn{K_{\mathrm{SS}} = Y_{\mathrm{SS}}(1 - h_2(Q_{X_A|X_\mathcal{B}}) - \max_{B_j\in\mathcal{B}} h_2(Q_{Z_A|Z_{B_j}})).}

Somewhat remarkably, it turns out that the $(n,n)$-threshold secret sharing scheme with GHZ states has exactly the same rate as the conference key agreement scheme! In hindsight however, this is unavoidable because of the necessity of a GHZ secret sharing scheme being an $(n,n)$-threshold scheme which removes optimisation over the trusted subset and enforces each complementary subset to be a single player.
\begin{Protocol}

\textbf{Alternative version of the $N$-BB84 protocol} \\The following protocol can be used to perform QSS and CKA by means of multi-partite entanglement, which we will denote as mQSS and mCKA, respectively. Common steps are not particularly marked, steps solely for mQSS are marked by a $*$, steps for mCKA by a \textasciicircum.
\begin{itemize}
    \item [\textit{1.\textasciicircum}] \textit{A key is shared between Alice and the Bobs determining the basis choice for every round, for key generation this is the $Z$ basis, the check bits are obtained from measurements in the $X$ basis.}
    \item [\textit{2.}] \textit{Alice distributes an $N-1$-entangled state to the participants over the bottleneck network via quantum network coding. This is equivalent to an $N$-partite being distributed to all players and Alice, as Alice can send the state according to her fictitious outcome. This is explicitly described in Appendix \ref{qber_calc}.}
    \item [\textit{3.}] \textit{The Bobs measure their respective particle of the multi-partite entangled state in $X$ or $Z$. In mCKA, the type of measurement is known beforehand (see \textit{1.\textasciicircum}), in mQSS, the key basis is chosen with probability $p_{\mathrm{key}}$.}
    \item [\textit{4.$*$}] \textit{All players announce their measurement bases in any order. Rounds, in which all parties measured in the $X$ basis can be used for key generation. Rounds, in which Alice and at least one Bob measured in $Z$ can be used for parameter estimation. This process is repeated until a sufficiently high number of secret bits is generated.}
    \item [\textit{5.}] \textit{Alice announces a random subset of check bits, for which the players also announce their measurement outcome. In mCKA, that is the collective $X$ measurement, in mQSS Alice makes the check independently with the Bobs in the $Z$ basis. For mCKA, Alice computes a correlation measure for the entire setup, i.e., for all Bobs. In mQSS, Alice computes a correlation measure with each Bob individually. If the correlation is sufficiently high the legitimate parties carry out error correction (sometime called information reconciliation), including a final hash-based correctness test. If either the initial correlation test or the error correction test fail, the protocol aborts.}
    \item [\textit{6.}] \textit{If the protocols does not abort, this will result in correlated strings between Alice and the Bobs. Finally, they perform privacy amplification to obtain final strings ($\mathbf{S}_A, \mathbf{S}_{B_i}$).}
    \label{Prot}
\end{itemize}
\end{Protocol}
\section{Bottleneck networks}
\begin{figure}[htb]
\includegraphics[width=0.26\textwidth]{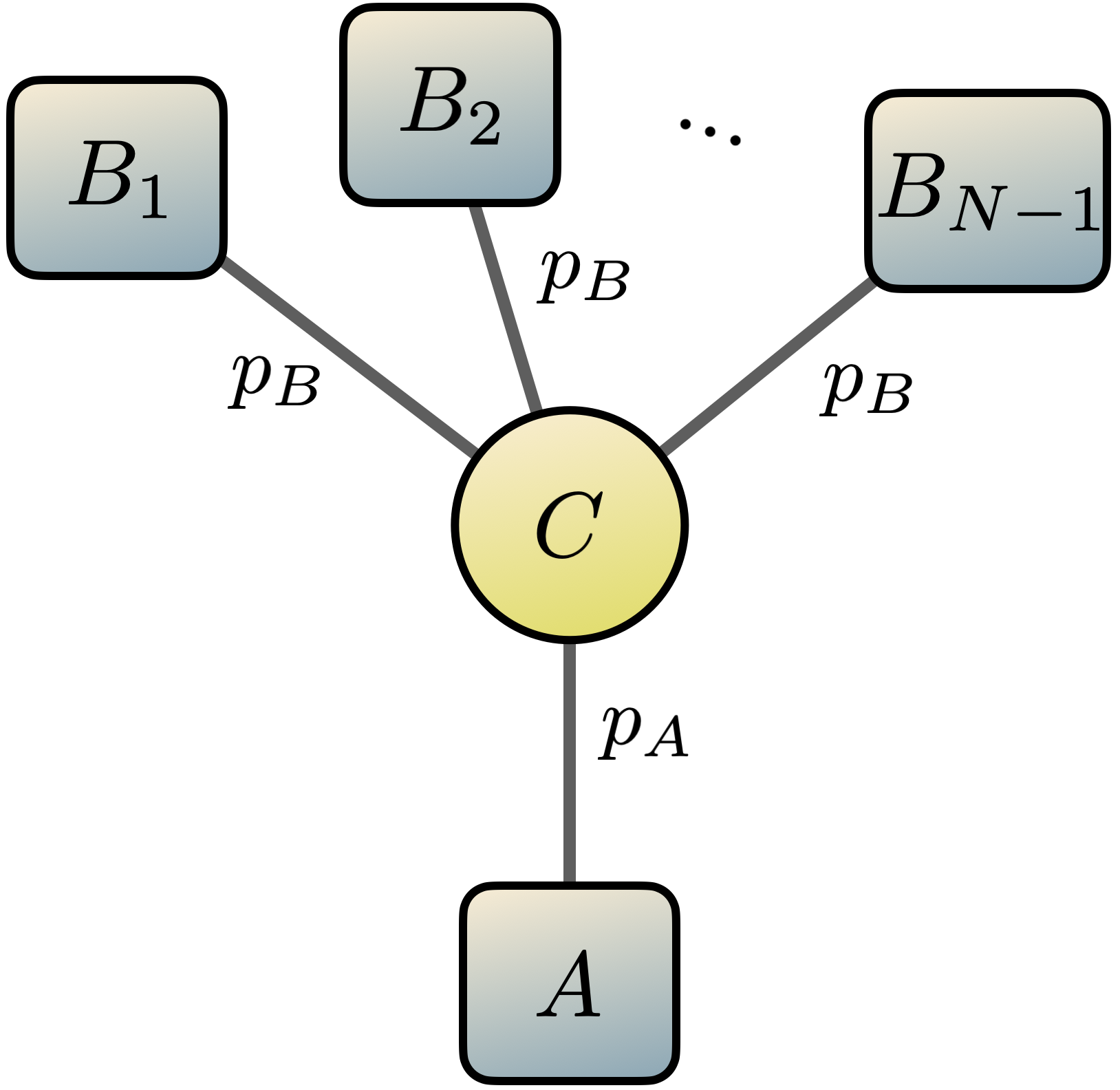}
\caption{Quantum cryptography in a network with a bottleneck. Alice is connected to a central router station $C$, which is connected to each of the Bobs by a quantum channel. $C$ must be able to produce and entangle qubits. Using the GHZ-based protocol,  where Alice sends a single qubit to $C$ which produces the multi-partite entangled state and sends one qubit to each Bob, only one single network use is required. However, when using the protocol based on bi-partite entanglement, the network has to be used $N-1$ times to establish the needed links. For a symmetric network, i.e., all links are of the same length, it holds that $p_A=p_B=p$. \label{network}}
\end{figure}

Recently it has been shown how GHZ states can be successfully distributed in quantum networks using quantum network coding \cite{Epping:2016hc, Hahn:2018wq}. 
We will focus on networks with so-called bottlenecks, as they can generate a performance speed up for protocols based on multi-partite entanglement via the number of network uses \cite{Epping:2017dx}. A bottleneck is a central station to which all involved parties are connected. It must be able to produce and entangle qubits to participate in the communication process (compare Fig.~\ref{network}). Using ordinary point-to-point QKD $N-1$ links must be established between Alice and every single Bob. This lower bounds the total number of network uses to be at least $N-1$. In contrast, protocols based on multi-partite entanglement distribute the photons to the Bobs by one single network use only. 

\subsection{Yields}

In the case of optical fiber the transmission probability $p$ scales exponentially with the distance $d$ (in km) as \eqn{p(d)=10^{-0.02d}.} This scaling behaviour is one of the biggest limitations on QKD protocol's performances, giving rise to the need for implementing quantum repeaters \cite{Luong:2016, Guha:2015}, aimed at overcoming 
limitations for quantum communication
\cite{Pirandola:2017jk,PhysRevResearch.4.023158,Wilde2017}. 
For multi-partite QSS several channels must successfully transmit photons simultaneously for a successful round. Ordinary point-to-point QKD only requires the $N-1$ links to work independently. For reasons of consistency, we will denote bi-partite QSS and CKA as bQSS and bCKA, respectively. In a completely symmetric network (compare Fig.~\ref{network} with $p_A=p_B=p$) this leads to the yields of bi-partite ($Y_{\mathrm{b}}$) and multi-partite protocols ($Y_{\mathrm{m}}$)\eqn{Y_{\mathrm{b}}=\eta \frac{p^2}{N-1}\, ,  \qquad Y_{\mathrm{m}}=\eta\, p^N.\label{yields_sym}}
$\eta>0$ is a factor taking into account the possible basis mismatch of the dealer and the players. Since both bi-partite CKA and QSS are regular QKD schemes, in which the parties can simply agree on a pre-shared key determining the measurement basis of each round, no rounds have to be discarded. Denoting to probability to measure in the key basis as $p_{\mathrm{key}}$, the probability for obtaining a round that can be used for key generation as $\eta_{\mathrm{k}}$ and the probability for obtaining a round that can be used for parameter estimation as $\eta_{\mathrm{c}}$ this means for bi-partite CKA and QSS that $\eta^{\mathrm{bCKA}}_{\mathrm{c}}=\eta^{\mathrm{bQSS}}_{\mathrm{c}}=1-p_{\mathrm{key}}$.
Further, this holds for multi-partite CKA as well, as all players are assumed to be trusted and a pre-shared key can again be used, \eqn{\eta^{\mathrm{mCKA}}_{\mathrm{k}}=p_{\mathrm{key}}, \hs \eta^{\mathrm{mCKA}}_{\mathrm{c}}=1-p_{\mathrm{key}}.\label{eta_key_check_CKA}} However, in multi-partite QSS, the situation is substantially different as the players are potentially dishonest and are not allowed to know the basis choice, and so the type of measurement, beforehand. So a valid key generation round will require all players simultaneously choosing the key basis and a valid check round requires Alice and at least one of the Bobs choosing the check basis. This leads to \cite{ourqss}
\eqn{\eta^{\mathrm{mQSS}}_{\mathrm{k}}=p_{\mathrm{key}}^N, \qquad \eta^{\mathrm{mQSS}}_{\mathrm{c}}=(1-p_{\mathrm{key}})(1-p_{\mathrm{key}}^{N-2}). \label{eta_key_check_SS}}
A round for key generation requires all parties to measure in the key basis, a round for parameter estimation requires the dealer and at least one player to measure in the check basis. With $L$ being the total number of rounds, we can express the rounds used for key generation $m$ as $m=\eta^{\mathrm{mQSS}}_{\mathrm{k}} L$ and the rounds for parameter estimation $s$ as $s=\eta^{\mathrm{mQSS}}_{\mathrm{c}} L$. 

While this acts as a penalty for the multi-partite protocol in finite size implementations, in the asymptotic limit ($L\rightarrow \infty$) $m$ and $s$ will tend to infinity for any $\eta^{\mathrm{mQSS}}_{\mathrm{k}}$ and $\eta^{\mathrm{mQSS}}_{\mathrm{c}}$. Consequently, in the asymptotic limit we can take $p_{\mathrm{key}}=1$ and so $\eta^{\mathrm{mQSS}}_{\mathrm{k}}=1$. 
The exponential scaling in the mQSS yield means optimizing the basis choice probability is especially important for performance with a large number of participants.

\subsection{Channel depolarization}

\begin{figure}[htb]
\includegraphics[width=0.53\textwidth]{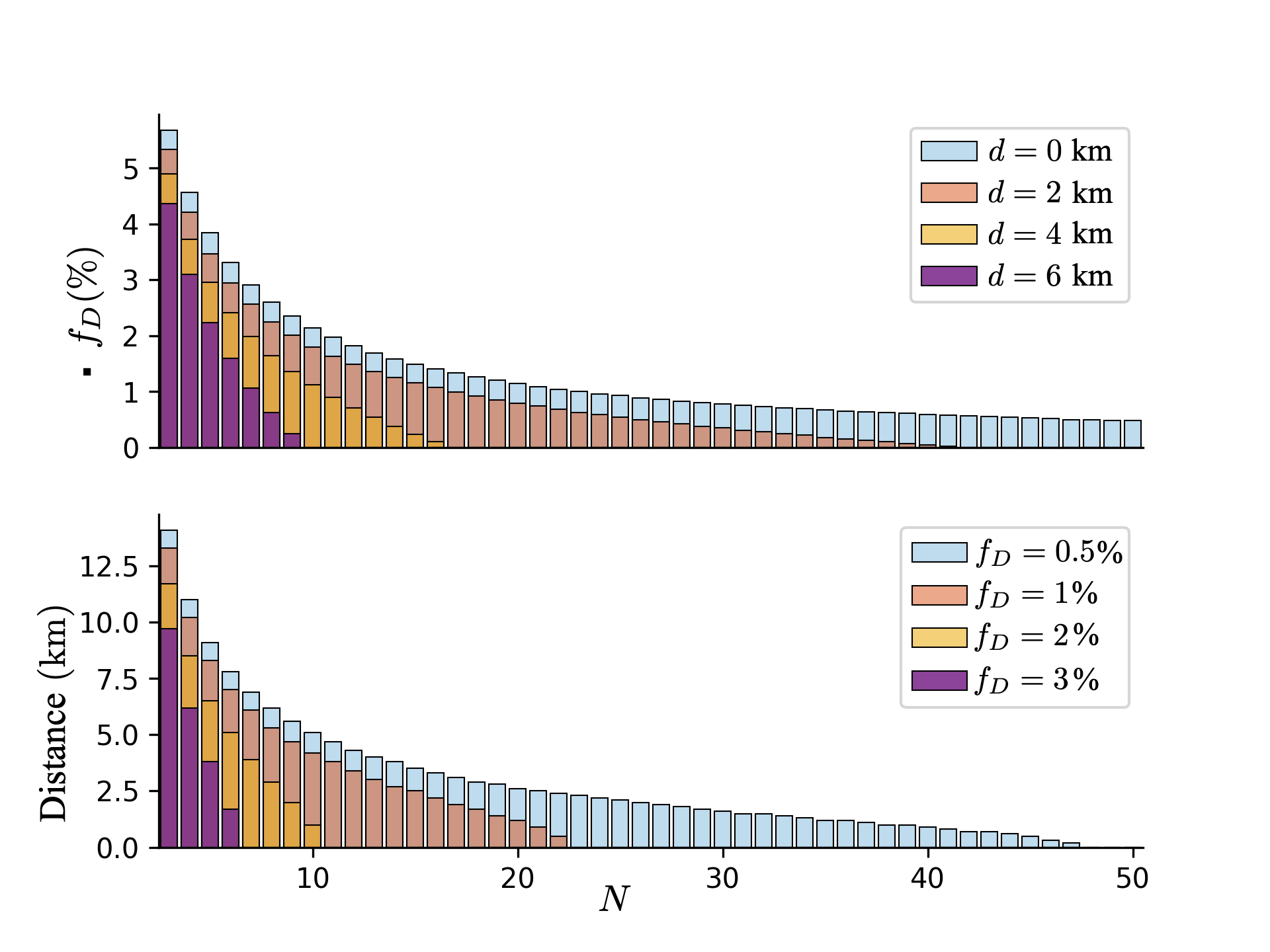}
\caption{Multi-partite advantage thresholds on the channel depolarization $f_D$ (top) and on different single link distances $d$ (bottom) of a symmetric bottleneck network. Below the threshold values the GHZ-based protocol is more efficient than its bi-partite competitor. Transmission losses of the network links have a stronger impact on the GHZ-protocol than on point-to-point QKD. For links longer than 15.1 km (corresponding to a transmission probability $p$ below 50\%), the performance advantage of the GHZ-based protocol completely vanishes. \label{threshold_channel}}
\end{figure}

In addition to transmission loss, which causes photons to simply not arrive at their designated destination, there is also noise acting on the channels blurring the signal and in that way affecting the amount of distillable key. We apply the noise model from Ref.~\cite{Epping:2017dx}, which they have used for their comparison of GHZ-based vs bi-partite protocol of CKA using the six-state-protocol. The QBER in the $Z$ basis between Alice and one single Bob $Q_{{A,B_i}}$ and the QBER for the collective $X$-measurement $Q_{\mathrm{X}}$ are \eqn{Q_{{X}}^{{\rm Ch}}=Q_{{A,B_i}}^{{\rm Ch}}=\frac{1}{2}(1-(1-f_{D})^N),}
where $f_D$ is the probability of depolarization of one channel. We proceed in a similar manner to Ref.~\cite{Epping:2017dx} and derive threshold values on the channel depolarization by numerically determining the intersection of bQSS and mQSS rates for a fixed number of participants $N$. The thresholds on the channel depolarization, for which mQSS still achieves higher rates than bQSS be found in Fig.~\ref{threshold_channel}. 
\begin{figure}[htb]
\includegraphics[width=0.53\textwidth]{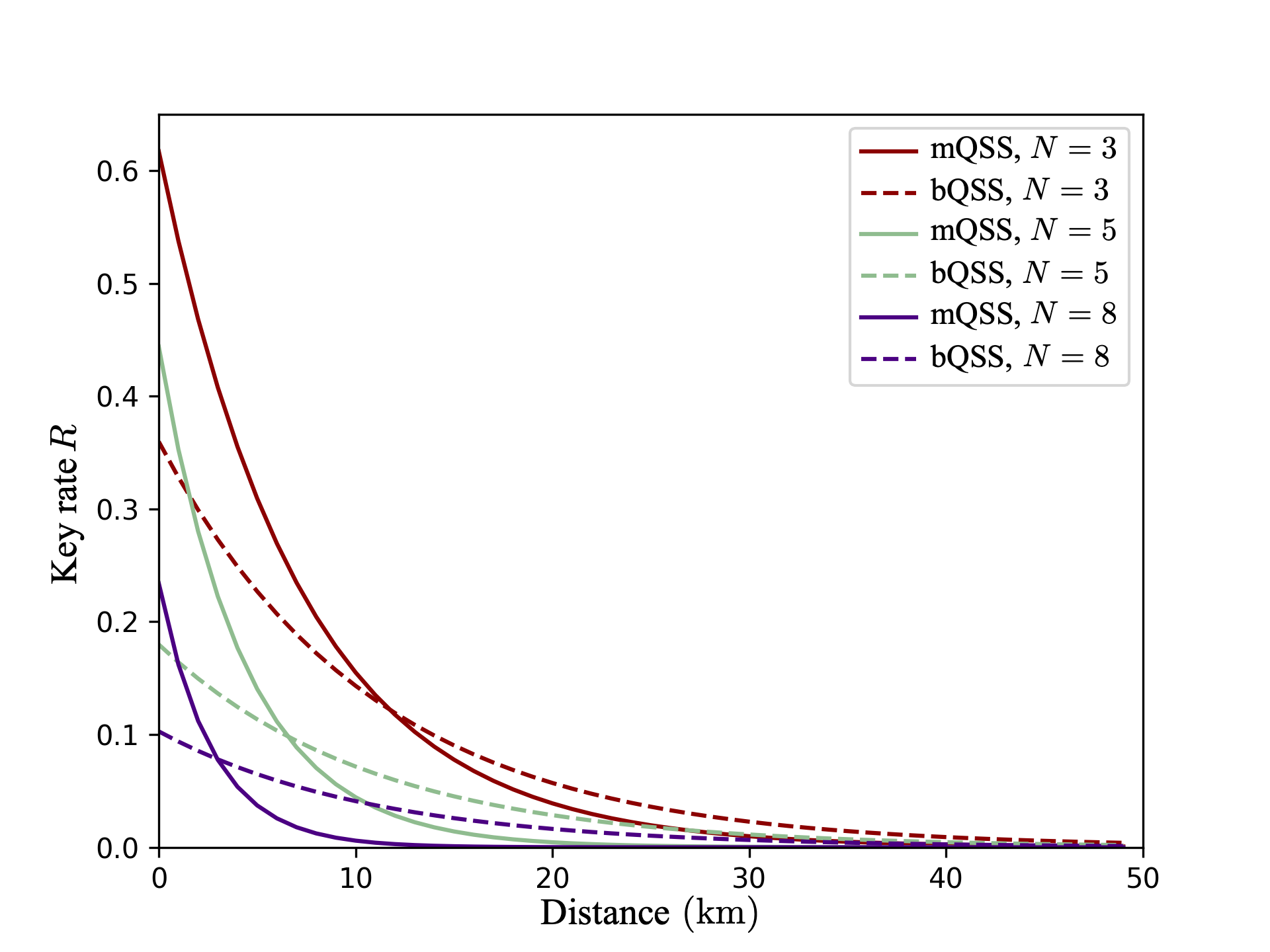}
\caption{Key rates for the GHZ-based secret sharing protocol (mQSS, solid lines) and its bi-partite competitor (bQSS, dashed lines) for a different number of participants, with a channel depolarization of 2 \%. It can be seen that the mQSS protocol is more sensitive to transmission loss. \label{depol_example}}
\end{figure}
For $d=0$ km, when Alice and the Bobs are hence not spatially separated at all, we recover similar thresholds to the ones from Ref.~\cite{Epping:2017dx}. The difference originates in the protocol used, which is the $N$-six-state protocol in their case in contrast to $N$-BB84. Adding transmission loss, which inevitably grows with the distances between Alice and the Bobs confirms the expected behaviour: it drives down the performance of all protocols, but degrades mQSS performance significantly more as in the multi-partite case several lossy channels have to work simultaneously. For distances of a single link of 15 km, i.e., a total distance from Alice to any Bob of 30 km, any performance advantage for the GHZ-based protocol vanishes. However, changing the network architecture towards an asymmetric network with one long link from Alice to the router station and shorter links connecting the router and the Bobs ($p_A\ll p_B$ in 
Fig.\ (\ref{network})) means an advantage can still be demonstrated for large distances between Alice and central node, as in such a network the short link's transmission govern the scaling of the advantage region. Such an architecture would be very natural in a setting where, for example, a single long link connects central hubs in different cities, which then distribute to multiple uses over a network of short links within a single metropolitan area.

In contrast to Ref.~\cite{Epping:2017dx} our analysis refers to the $N$-BB84 protocol, making the results applicable to both CKA and secret sharing as shown in Section \ref{secret_sharing_rates}. Note here that the usage of a pre-shared key determining the basis choice, which is only allowed for CKA as it would undermine the security of a secret sharing protocol, will only have an impact in a finite-size-analysis. In the asymptotic limit, it will not. 

It would also be interesting to expand this comparison to $N$-six-state secret sharing, although preliminary considerations suggest a multi-partite advantage would be unlikely in this case. The nature of the GHZ-state forbids using collective $X$-measurement for parameter estimation, as it would mean dishonest agents could manipulate which rounds discarded and re-open the participant attack. The $Z$-measurement, however, 
cannot be used for key generation as it immediately fails to satisfy the desired secret sharing access structure. Using valid combinations of $X$- and $Y$-measurements, as proposed in the original protocol of Hillery, Berthiaume and Bu{\v z}ek \cite{Hillery:1999tb} would force to sacrifice a lot more data for parameter estimation by the additional classical post processing  \cite{Williams:2019kb}. For every measurement basis $N-1$ sets have to be recorded with each Bob going first once. This would always act as a penalty for mQSS and means in this setting bQSS will always achieve higher rates and be more efficient when all three measurement bases are used.

\section{Memory network}\label{memorynetwork}

Equipping the central station and the participants with \emph{quantum memories} (QM) can increase the advantage region for the mQSS protocol by bypassing that need for all links to simultaneously succeed. We suggest the following addition to the protocol: The central station prepares $N-1$ Bell pairs $\ket{\phi^+}$, keeps one photon of each pair by storing it in a quantum memory and sends the other half to each Bob $B_i$, respectively. As soon as one of Alice's photons successfully arrived at the central station, the GHZ-state is prepared and sent to each one of the Bobs by entanglement swapping. The rest of the protocol, the measurements of the Bobs for key generation or to obtain the check bits, is similar to before. 

\subsection{Yields}

With $p_A$ being the transmission of the long, lossy link and $p_B$ the transmission of the shorter links connecting the central station and the Bobs, the yields for both schemes are
\eqn{Y_{\mathrm{b}}=\eta \min \Big\{\frac{p_A}{N-1},p_B\Big\},  \qquad Y_{\mathrm{m}}=\eta \min \Big\{p_A,p_B\Big\}\label{yields_asym}.}
However, as we have chosen  $p_A\ll p_B$, 
by definition whenever the long link has succeeded, the shorter link will have succeeded too. This means (\ref{yields_asym}) can be evaluated to 
\eqn{Y_{\mathrm{b}}=\eta \frac{p_A}{N-1},  \qquad Y_{\mathrm{m}}=\eta\, p_A.\label{yields_asymII}}
Effectively, this is the same scaling of the mQSS advantage region as the case without loss. 
However, it should be kept in mind that yields are still scaling with the transmission probability of the longer link and only the ratio of both is the same.

\subsection{Channel depolarization}

The depolarization of a single subsystems $i$ 
of a multi-partite state will be modelled in the following way \cite{nielsen_chuang:2010}
\eqn{
\mathcal{E}_i(\rho)=(1-\frac{3f_{\mathrm{D}}}{4})\rho+\frac{f_{\mathrm{D}}}{4}(X_i\rho X_i+Y_i\rho Y_i+Z_i\rho Z_i), 
\label{dep_channel}}
where $f_D$ is the probability of depolarization. Note here that the depolarization, in contrast to the transmission loss, is not dependent on the length of the channel.

\subsection{Quantum memories}
We will use the following map to adequately model the dephasing of subsystem $i$ after an elapsed time interval $t$ \cite{Razavi:2009}
\eqn{
\Gamma_i(\rho)=(1-\lambda_{\mathrm{dp}}(t)) \rho + \lambda_{\mathrm{dp}}(t) Z_i \rho Z_i.
\label{deph_map}}
$Z_i$ is the Pauli $Z$ operator acting on $i$, the dephasing coefficient $\lambda$ is time-dependent as
\eqn{
\lambda_{\mathrm{dp}}(t)=\frac{1-e^{-t/T_2}}{2},
\label{dephasing_coefficient}}
where $T_2$ is the device-dependent dephasing time of the QMs. In the proposed addition to the protocol several qubits of an entangled state dephase simlutaneously, which simply is a concatenation of the map from (\ref{deph_map}). 

\subsubsection{Dephasing intervals}
The time that elapses until a photon emitted from Alice arrives at the central station is given by
\eqn{
\tau_A = T_p + \frac{d_A}{c} \label{tau_A},}
where $T_p$ is the preparation time for a bi-partite entangled state, $d_A$ the distance between Alice and the central station and $c$ the speed of light in optical fiber ($2\times 10^8 \frac{\text{m}}{\text{s}}$). To establish the pre-shared Bell pairs between the central station and the Bobs, the central station will produce a Bell pair for each Bob, store one photon in the local QM and send the other one through the quantum channel to the corresponding Bob. Subsequently, the central station will await a signal from the respective Bob confirming the photon's arrival. Throughout this paragraph, subscripts for the Bobs and hub sites are omitted, given the identical setups. The trial time required to establish one pair is then given by
\eqn{\tau_{B}= T_p + \frac{2d_{B}}{c} \label{tauB_i}.}
Note here that the distribution of the Bell pairs to the Bobs takes place simultaneously. Upon the successful arrival of a photon from Alice at the central station, a GHZ state is produced via quantum network coding (see Appendix \ref{qber_calc}). Subsequently, the state is swapped to the end nodes $B_1, \dots, B_{N-1}$ via the pre-established Bell pairs. We assume here that the execution of the gates for the GHZ state production and entanglement swapping measurement is fast relative to transmission times and can be neglected. The average waiting time for a photon at each Bob's site can be expressed as
\eqn{
t_{B}=N_A\tau_{A}-N_{B}\tau_{B}+\frac{2d_B}{c}, \label{tB_i}}
where the last terms accounts for the classical communication, first needed to signal the successful photon reception of the photon of the Bell pair and later to communicate the needed correction gate depending on the outcome of the entanglement swapping measurement. The average waiting time for the photon of the Bell pair stored at the hub $\tilde C$ is identical,
\eqn{
t_{\tilde C}=N_A\tau_{A}-N_{B}\tau_{B}+\frac{2d_B}{c}, \label{tC_i}}
however here the last term originates from the fact that the qubit at the hub is already in the QM when the paired qubit travels to the remote site (Bob), and the confirmation for successful transmission occurs. Upon the arrival of Alice's qubit at the central station, the qubit $\tilde C$ can be immediately measured. \\
Ultimately, we are interested in the expected dephasing for each individual Bob and at the central station, i.e., the values of $e^{-t_{B_i}/T_2}$ and $e^{-t_{\tilde{C}_i}/T_2}$, to determine the QBER. However, in addition to the point in time when Alice's photon is detected, this dephasing depends on the round in which the last Bob successfully establishes its link and the round in which the specific Bob successfully detects a photon. 
All $N_{B_i}$ are symmetrically distributed around a certain expected value, as are the different combinations of all $N_{B_i}$. However, including the highly non-linear dephasing, the true value of the expected dephasing deviates from the value obtained by computing it independently for each $N_{B_i}$, making the overall analysis significantly more complex. For a fully symmetric setup, this problem has been considered analytically in Ref.~\cite{Avis:2023jy}. Given that we consider a slightly asymmetric variation, we evaluate the expected dephasing by numerically sampling the random variables $N_A$ and $N_{B_i}$ for a given sample size.

\subsection{Key rate analysis}

\begin{figure}[htb]
\includegraphics[width=0.49\textwidth]{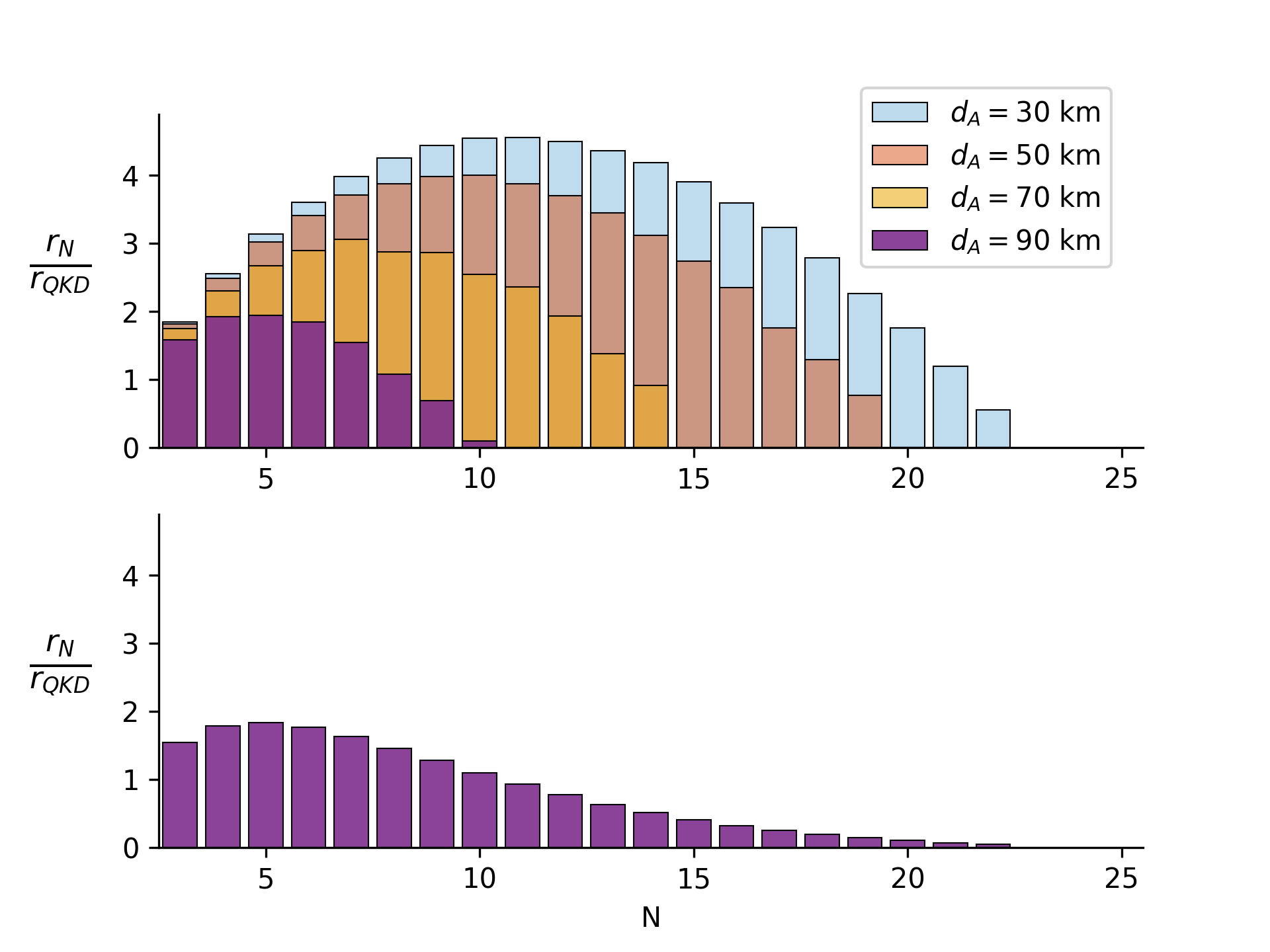}
\caption{Performance comparison of mQSS to bi-partite rates with (top row) and without (bottom row) quantum memories in an asymmetric network. In the top row, we also allow the bipartite protocol to optionally use a quantum memory and compare to the optimal bipartite rate. Memory dephasing is modeled with an internal dephasing time of the QMs as $1$\,s and a preparation time of a bi-partite entangled state of $2$\,\textmu s. The expected dephasing (see Equation~(\ref{expected_dephasing_1}) and (\ref{expected_dephasing_2})) needed for calculated the QBERS is numerically evaluated over a sample size of $10^3$. In both rows the channel depolarization is fixed to $0.01$ and the short link's transmission to $d_B=4$ km. The long link's transmission $d_A$ cancels out in the comparison without QMs. Crucially, the advantage for GHZ-based protocols in the network utilising QMs increases up to a much larger $N$ than in former investigated network configurations. \label{memrates}}
\end{figure}

We now have all the ingredients to calculate quantum bit error rates for both schemes to evaluate the key rates. To carry out a fair comparison, we will also equip the bi-partite protocol with QMs. The states will be distributed over the bottleneck network according to the network coding scheme of Ref.~\cite{Epping:2016hc}. Detailed calculations yielding the 
\emph{quantum bit error rates} (QBER) can be found in Appendix \ref{qber_calc}. The evaluated key rates for mQSS and bQSS are shown in Fig.~\ref{memrates}. The internal dephasing time $T_2$ is chosen to be $1$ s, in
Ref.~\cite{Olmschenk2007}, a dephasing time of $2.5$ s has been observed. The preparation time for a bi-partite entangled state $T_p$ has been fixed to be $2\,\mu$s as in Ref.~\cite{Luong:2016}. Fig.~\ref{memrates} shows the ratio of multi-partite and bi-partite rates in an asymmetric network with or without QMs. 

Note here that in the top row, we always compare to the maximum of bi-partite rates with or without QMs. We observe a linearly increasing advantage up to some number $N$ that depends on the distance of the long link $d_A$ ($N\approx11$ for $d_A=30$ km). The same parameters only give a linear advantage up to $N\approx5$ in the memoryless case. The total advantage of the mQSS protocol using QMs over bQSS (i.e., when the ratio of both is bigger than 1) up to a number of participants of $N=20$ for $d_A=30$ km, whereas this is $N=10$ in the memoryless case. Note here that we are indeed in a low loss regime in terms of depolarizing noise, which may deviate from experimental implementations given that depolarizing noise of imperfect entanglement swapping measurements augments exponentially in participant number. 

We have convincingly shown that protocols based on multi-partite entanglement in networks with bottlenecks can indeed beat the benchmark of ordinary point-to-point QKD. Whereas this shrinks when we go to larger distances, employing QMs unlocks a performance advantage for much further distances than previously. This advantage remains significant when including proper modeling of memory dephasing with realistic parameters.

\section{Finite-size analysis}

To complement this in-principle-advantage, and to relate the findings to practical applications, in this section, we provide a composably secure analysis and investigate the finite-size performance of multi-partite entanglement based schemes with and without memories. The \emph{composable security framework}  \cite{Renner:2005p464,Tomamichel:2012p7120,Portmann:2022io}
has been extended to the case of secret sharing in Ref.~\cite{ourqss} and we will also make extensive use of the analysis of multi-partite CKA carried out in Ref.~\cite{Grasselli:2018gk}. Ref.~\cite{Grasselli:2018gk} has extended earlier work \cite{Tomamichel:2012p7120} that leveraged entropic uncertainty relations \cite{Tomamichel:2011p461} to provide a finite-size, $\varepsilon$-secure proof for CKA and the corresponding result for secret sharing (for the case of discrete and continuous variable case based upon earlier asymptotically analyses \cite{Kogias:2017jz,Walk:2016jx}) has been derived in Ref.~\cite{ourqss}. Here, we present the key results, with detailed proofs and definitions deferred to Appendix \ref{finite}. Roughly speaking, the goal of such an analysis is to consider a protocol involving a finite number of rounds $L$, and calculate a finite string length $\ell$, and an $\varepsilon>0$ which can be meaningfully regarded as quantifying the deviation from the output of an ideal protocol. 

\begin{figure}[htb]
\includegraphics[width=0.49\textwidth]{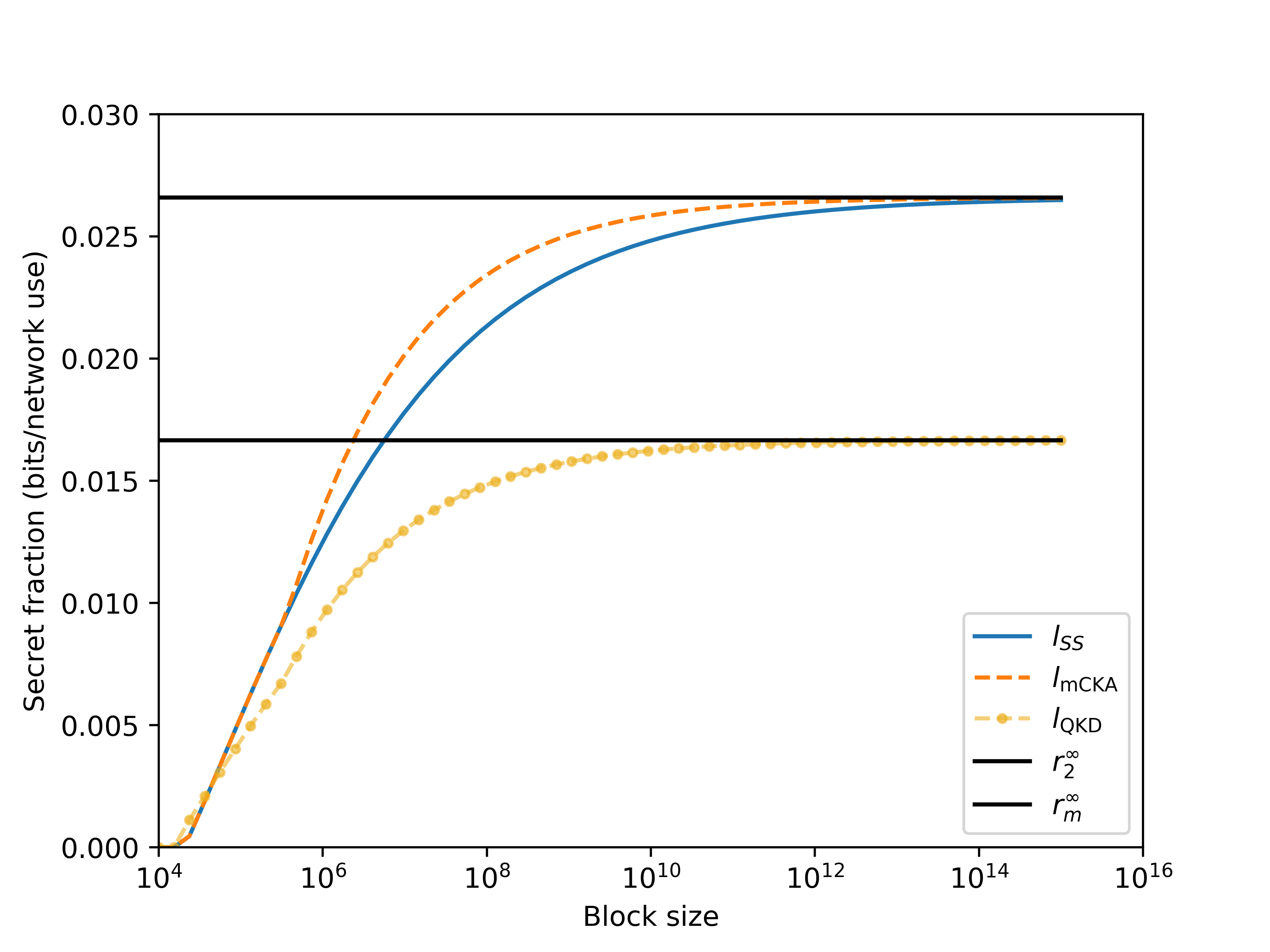}
\caption{Secret fraction as a function of block size for multi-partite entanglement based QSS (solid blue) protocol where active basis switching is essential, CKA (dashed red) and bi-partite-QKD based rate protocol for either QSS or CKA. 
A pre-shared key maybe used instead of active basis switching for CKA and bi-partite protocols. Performance in the 
asymptotic limit for multi-partite and bi-partite protocols (solid black) 
are plotted for comparison. 
Network parameters are $d_A = 50$ km, $d_B = 4$ km, $f_D = 0.01$ and the security parameter is $\varepsilon = 10^{-10}$. The basis probability 
is numerically optimised at each point.\label{fig:symm_blocksize}}
\end{figure}

More concretely, define $\rho_{\bm{S}_A,\bm{S}_\mathcal{B},E}$ as the joint state of the final strings (after all post-processing) generated by Alice, the set of Bobs and the 
eavesdropper conditioned on the protocol not aborting. The parameter $\varepsilon$ is an upper bound to the joint probability of the protocol passing all tests and this conditional state being successfully distinguished from the output of an ideal protocol. In that sense, it can be interpreted as a failure probability in that it is the probability of undetected imperfection, i.e., all of the tests are passed while the output is in some way `bad'. Typically, this parameter is decomposed into a correctness parameter, $\varepsilon_c$ (the joint probability of passing the test and finding the strings of the legitimate parties are not appropriately correlated), and a secrecy parameter, $\varepsilon_s$ (the joint probability of passing the test and the conditional joint state being distinguishable from a uniform distribution in a tensor product with the malicious parties). We refer to a protocol that has provable bounds on these quantities as being $\varepsilon$-secure with $\varepsilon = \varepsilon_s+\varepsilon_c$. 

The first significant point is that, in the finite-size regime, we cannot simply take the limit where the check basis probability tends to zero and optimising over this parameter becomes non-trivial. The different yields, given by Eq.~(\ref{eta_key_check_CKA}) and Eq.~(\ref{eta_key_check_SS}), 
reflect the possibility for CKA or QKD protocols to utilise pre-shared key as all participants in each protocol are trusted. This means the performance of QSS and CKA protocols are no longer necessarily identical over a given network. In Fig.~\ref{fig:symm_blocksize} we plot a representative example of this effect. We see that, whilst the both protocols eventually asymptote to the same value for sufficiently large block sizes, there are regions where the CKA protocol, by exploiting the ability to use pre-shared key, is able to achieve better performance. Interestingly, whilst it is intuitive that both strategies should coincide in the large block-size limit, our analysis also indicates that, for sufficiently small block-sizes, active basis switching is optimal and so the CKA and QSS protocols again coincide. Finally, we note that in this figure and all others, when comparing to bi-partite protocols we always take the maximum over both strategies for the bi-partite case. Further discussion and results, including the optimal basis probabilities can be found in Appendix~\ref{App:extra_results}. 

\begin{figure}[htb]
\includegraphics[width=0.5\textwidth]{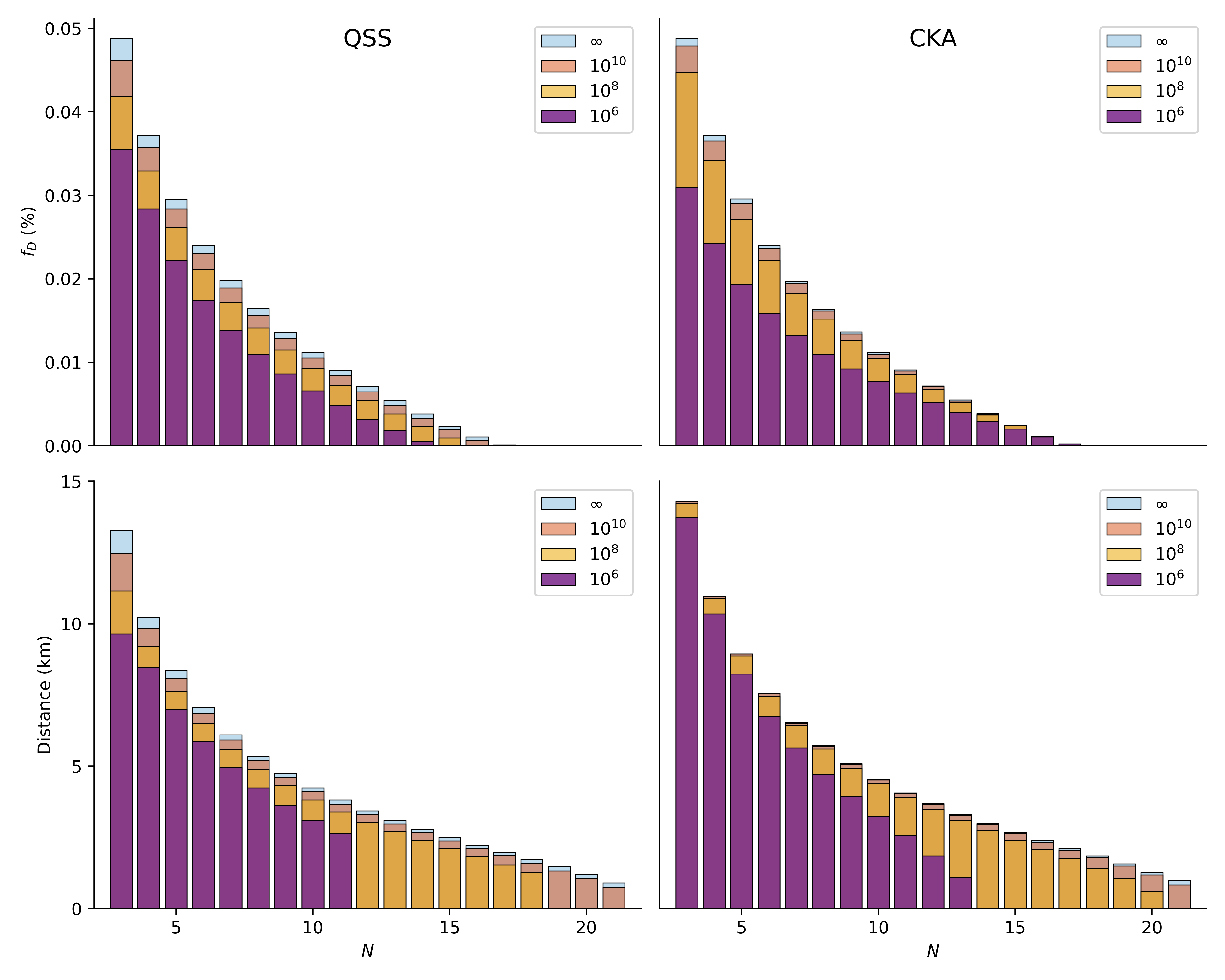}
\caption{Multi-partite advantage thresholds for a symmetric bottleneck network in terms of depolarising channel noise $f_D$ (top row) and transmission distance (bottom row) for quantum secret sharing (left column) and conference key agreement (right column) for various block sizes. In the top row the transmission distance is fixed to 4 km and in the bottom row the noise is fixed to 1\%.\label{fig:symm_finite_thresholds}}
\end{figure}

In Fig.~\ref{fig:symm_finite_thresholds}, we return to the thresholds for multi-partite advantage in light of these finite-size issues. Here we will quantify the finite size of a given protocol by specifying a block size, $m$, of the number of detections in the key-generating basis and compare the multi-partite and bi-partite strategies for a given channel at the same fixed block size. For both QSS and CKA protocols without memories over a symmetric network, we plot depolarising noise thresholds for multi-partite advantage as a function of player number, first for depolarising noise for a fixed transmission distance of 4km and then for transmission distance with a fixed noise of $1\%$. For all cases we see that for a block size of $m=10^{10}$ the asymptotic performance is almost completely recovered. Naturally performance is curtailed for smaller block sizes, however a noise-tolerant advantage for networks of up to 10 players can be achieved for block sizes as small as $m=10^6$. The finite-size reduction in performance in terms of loss tolerance is much more severe than for depolarising noise, and the multi-partite advantage for CKA is slightly more robust than for QSS as expected given the additional trust assumptions.

\begin{figure}[htb]
\includegraphics[width=0.49\textwidth]{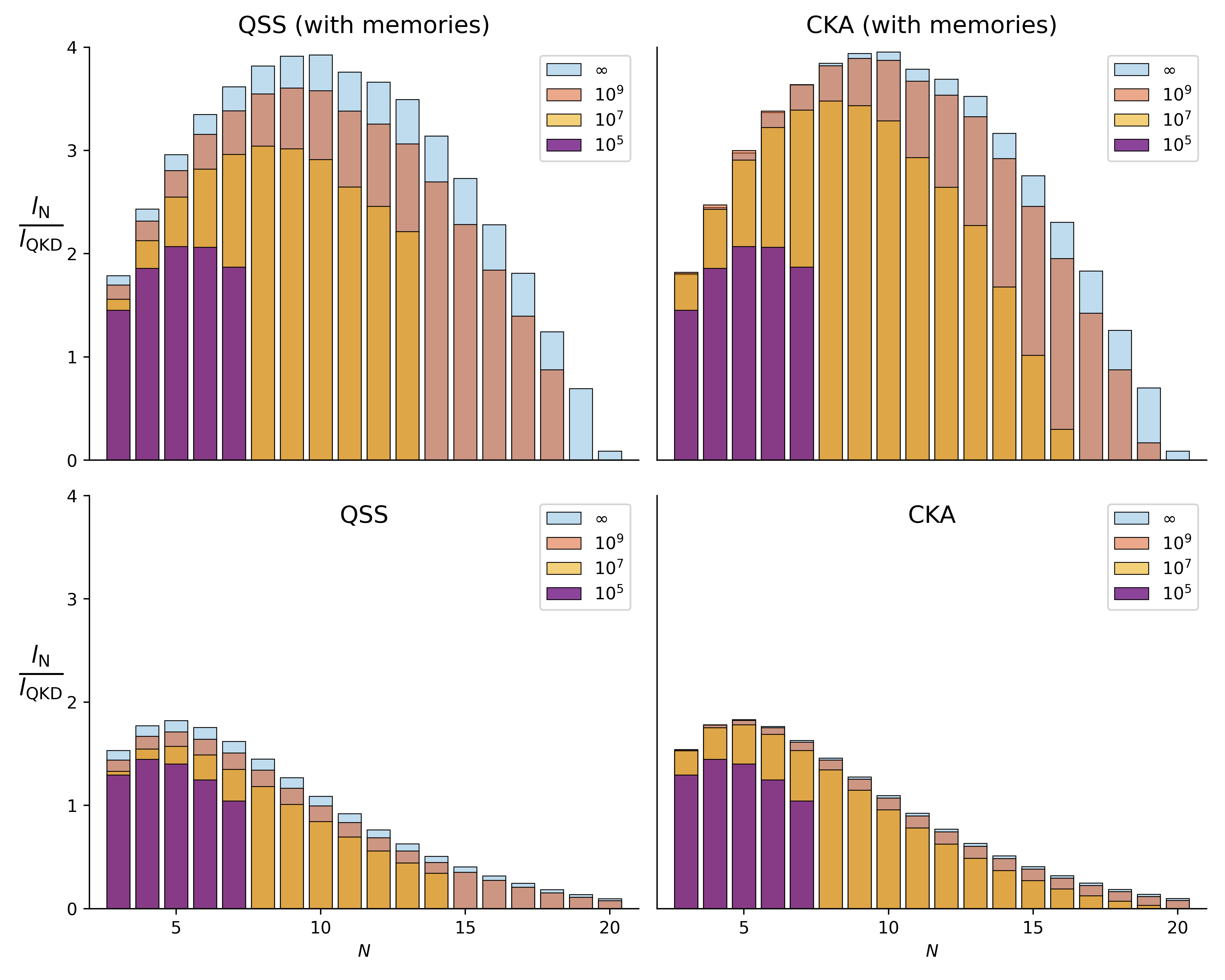}
\caption{Finite-size performance comparison of mQSS and mCKA as a function of player number with (top row) and without (bottom row) quantum memories. Network parameters are $d_A = 50$ km, $d_B = 4$ km and $f_D = 0.01$, the memory dephasing time is $T_2 = 1 $s, the preparation time of a bi-partite entangled state is $T_p = 2$ \textmu s. The expected dephasing has  again been numerically evaluated over a sample size of $10^3$. The security parameter is $\varepsilon = 10^{-10}$ and the basis probability is numerically optimised at each point. \label{fig:asymm_blocksize_memories}}
\end{figure}

Lastly, in Fig.~\ref{fig:asymm_blocksize_memories}, we consider the use of quantum memories to restore a scalable multi-partite advantage. When comparing to a QKD protocol, we always compare with the optimal QKD based strategy, i.e., optimised over whether active basis switching or pre-shared key is used as well as whether utilising memories is even advantageous. Crucially, we find that the use of quantum memories unlocks a multi-partite advantage that increases linearly in player number up to substantially larger networks than the memoryless case. For a reasonable asymmetric network where with a long link of $d_A = 50$ km, a short link of $d_B = 4$ km and depolarising channel noise of $1 \%$ we find that a quantum memory with a dephasing time of 1 s can achieve a linear advantage in participant number up until approximately 10 parties. Moreover, some multi-partite advantage can be observed up until 17 parties. This represents a substantial improvement over the case without memories where the improvement increases only until around 4 players and vanishes altogether after 7 players.

\section{Conclusions and outlook}

In this work, we have analysed an alternative variant to the classic HBB protocol for QSS based on multi-partite entanglement that is invulnerable to the participant attack. Unlike already known protocols for this purpose, our protocol does not require additional data to be sacrificed for parameter estimation. It fixes $X$ to be the key basis and $Z$ to be the check basis and proves to be extremely flexible for actual experimental implementations: swapping the designations of the bases allows an identical setup to perform CKA instead. 

We have shown that this multi-partite protocol has an distinct advantage over ordinary point-to-point QKD when evaluated over bottleneck networks. This advantage is also maintained when transmission losses, the major limitation for experimental quantum cryptography, and finite-size effects are taken into account. On the one hand, we find that the advantage regime, in terms of both player number $N$ and transmission distance, is limited to small networks. However, we further proposed to add quantum memories which generates an advantage for the multi-partite protocol even in the high transmission loss regime. 

Contrary to known GHZ-based protocols, the advantage for the multi-partite protocol here \emph{grows} with an increasing number of participants up until some finite number that depends upon the decoherence of the quantum network and the quantum memories. As well as providing a compelling example of the potential of multi-partite entanglement and its importance to when studying quantum cryptography in the future, we hope these results will also be useful to provide operational benchmarks for quantum network resources that simultaneously consider the quality of multi-partite entanglement generation and quantum repeater technology.

Numerous extensions and further research directions based on this work could prove fruitful. In the first instance, here we have considered ($N,N$) secret sharing, however several results are known regarding secret sharing for various ($N,k$)-schemes in terms of achievable access structure (without resolving the participant attack) \cite{Markham:2008ta,Keet:2010cf,Marin:2013dc} . A natural open problem would be to ascertain precisely if and when our security proof can be extended to these other instances. It would also be important to revisit previous results on continuous variable secret sharing \cite{ourqss} to examine to what extent analogous multi-partite improvement could be found based upon quantum memories in this regime. Finally, perhaps the most practically relevant open problem would be a systematic exploration of the optimal use of multi-partite entanglement in repeater networks, for both QSS and CKA. We were able to obtain analytic results in the limit of highly asymmetric networks and analytic progress on highly symmetric star networks has also recently appeared \cite{Avis:2023jy}. 

For general networks, it will presumably be necessary to resort 
to sophisticated numerical simulation tools for repeater architectures \cite{Wallnofer:2022ek,NetQASM,NetSquid,satoh2021quisp}. This would facilitate a systematic comparison of memory platforms for quantum cryptography along the lines of Ref.\ \cite{vanLoock:2020gt} as well as an analysis of optimal strategies (e.g., memory cutoff times or use of entanglement purification) in the multi-partite setting.

In this work, we have presented a ``smoking gun'' that signifies a robust multi-partite advantage beyond bi-partite  architectures. We do so for meaningful quantum cryptographic tasks of practical relevance. This work hence supports the line of thought that it is highly fruitful to explore quantum communication tasks beyond point-to-point architectures. It is the hope that this work stimulates more such efforts that relate to practical 
multi-partite communication tasks.
\section*{Acknowledgements}
The authors thank J. Walln{\"o}fer for helpful discussions and comments. JM thanks the Q-net-Q Project, which has received funding from the European Union’s Digital Europe Program under grant agreement No 101091732, and is co-funded by the German Federal Ministry of Education and Research (BMBF). Furthermore, JE and NW thank the BMBF (QR.X, PhoQuant, QPIC-1), the ERC (DebuQC), and the Einstein Foundation (Einstein Research on Quantum Devices) for funding.
\newpage 

\appendix

\section{QBER calculation including quantum memories}\label{qber_calc}
\begin{figure}[htb]
\includegraphics[width=0.27\textwidth]{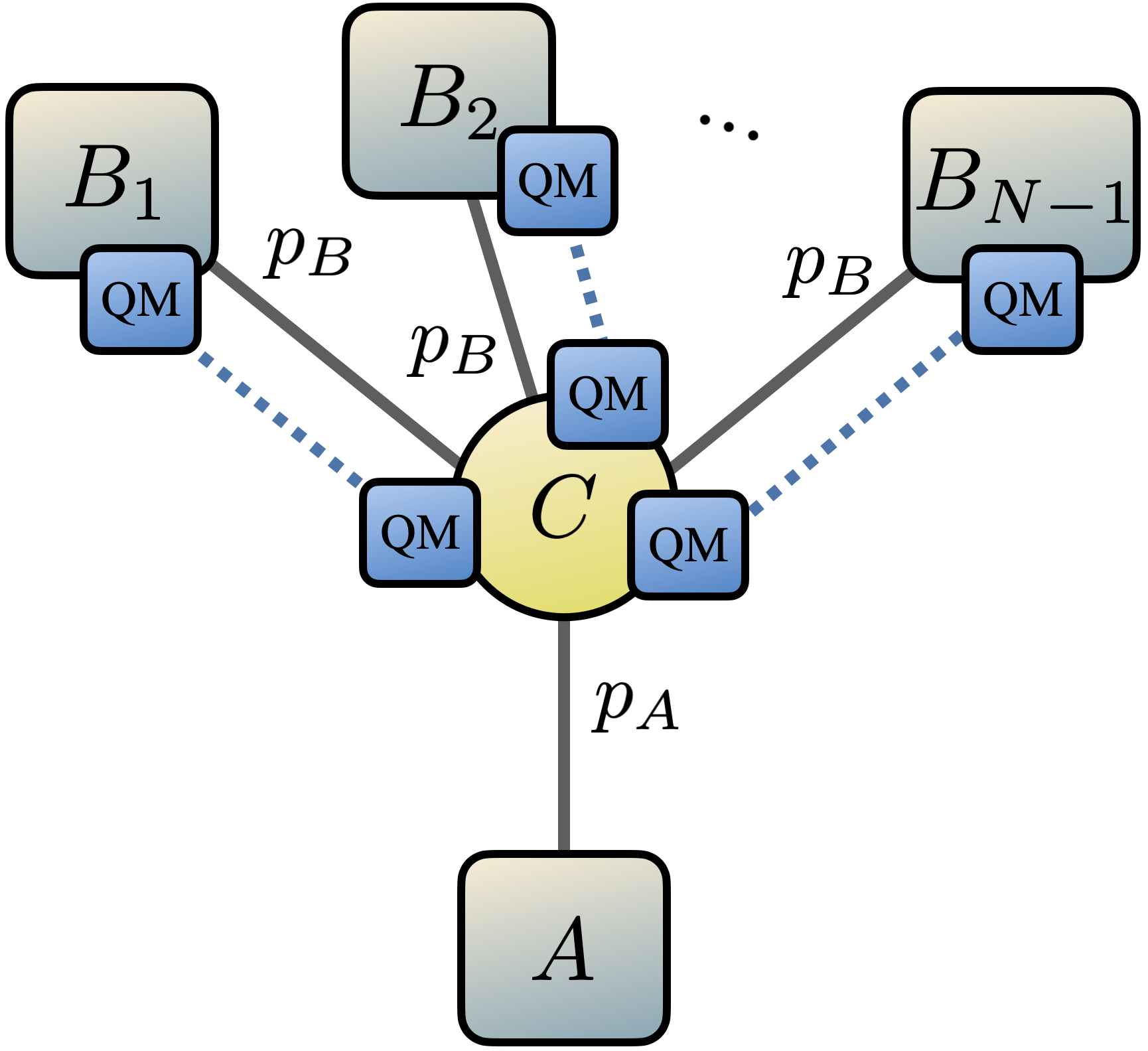}
\caption{A quantum bottleneck network using quantum memories (blue circles) at the hub and each of the Bobs' sites. These memories are used to store photons of a Bell pair shared between the hub and each of the Bobs. Once a photon from Alice travelling over the long link with transmission probability $p_{\mathrm{A}}$ is successfully detected at C, the multi-partite state is produced and projected to the Bobs via the Bell pairs. This is done by a Bell state measurement on the respective qubit of the GHZ state and the one of the Bell pair.\label{network_mem}}
\end{figure}

We will now give the explicit calculation to obtain the QBER expressions for the network including the use of quantum memories (see Fig.~\ref{network_mem}). Note that the QBER expressions for the bipartite protocol including quantum memories emerge from the multipartite expressions by setting $N=2$. We will follow the network coding scheme of Ref.~\cite{Epping:2016hc} to distribute a multi-partite entangled state in the bottleneck network, but in the computational basis. Alice produces two qubits $C$ and $A$ in the state vector $\ket{0}$ and applies a controlled-$X$ gate acting as 
\begin{equation}
C_X=\ket{+}\bra{+}\otimes\mathbf{1}+\ket{-}\bra{-}\otimes X 
\end{equation}
on them. At this point, the shared state vector  is of the form 
\eqn{
\ket{\psi}_{C,A}=\frac{1}{\sqrt{2}}(\ket{+,0}+\ket{-,1}).} She keeps photon $A$ and sends $C$ through the quantum channel to the router station, which depolarizes according to Eq.~(\ref{dep_channel}). In the case of the bipartite protocol, another qubit $C_i$ in the state vector $\ket{0}$ is produced and entangled with the remaining state by a $C_X$ gate acting $A$ and $C_i$. In the multi-partite case, $N-1$ qubits in the state vector $\ket{0}$ are produced and entangled with the qubit $C$ by $(N-1)$ $C_X$ gates. Note here that in the multi-partite case, the state would hold more qubits 
($C_1 ,\dots , C_{N-1}$), however, they would all be in the same state as $C_i$. The state at this point reads

\begin{align}
\rho_{C,A,C_i}=\frac{1}{2}\Big(1-\frac{3f_{\mathrm{D}}}{4}\Big)\Bigg(\Big(\ket{+,0,0}+\ket{-,1,1}\Big)&\Big(\dots \Big)\Bigg)\nonumber\\
+\frac{1}{2}\frac{f_{\mathrm{D}}}{4}\Bigg(\Big(\ket{+,1,1}+\ket{-,0,0}\Big)&\Big(\dots \Big)\nonumber\\
+\Big(\ket{+,1,1}-\ket{-,0,0}\Big)&\Big(\dots \Big)\nonumber\\
+\Big(\ket{+,0,0}-\ket{-,1,1}\Big)&\Big(\dots \Big)\Bigg),
\end{align}
where the brackets indicate the respective complex conjugate transpose.  
Qubit $C$ is then measured in the $Z$-basis, and depending on the post measurement state a correcting gate is needed. For the post measurement state vector $\ket{1}_C$ this is
$Z_{C_i}$ on one single qubit $C_i$, for the post measurement state vector $\ket{0}_C$ no correction is needed. The remaining qubits, depending on which protocol is used either $N-1$ or one single qubit $C_i$, are in the mixed state
\begin{align}
\rho_{A,C_i}=\frac{1}{2}\Big(1-\frac{3f_{\mathrm{D}}}{4}\Big)\Bigg(\Big(\ket{0,0}+\ket{1,1}\Big)\Big(\dots \Big)&\Bigg)\nonumber\\
+ \frac{1}{2}\frac{f_{\mathrm{D}}}{4}\Bigg(\Big(\ket{0,0}-\ket{1,1}\Big)\Big(\dots \Big)&\nonumber\\
+\Big(\ket{0,1}-\ket{1,0}\Big)\Big(\dots \Big)&\nonumber\\
+\Big(\ket{0,1}+\ket{1,0}\Big)\Big(\dots \Big)&\Bigg).
\label{rhoACi}
\end{align}

The repeater part of the protocol requires this state to be swapped to the receiving parties by means of pre-established Bell pairs shared by the router station and the Bobs. We apply the error models introduced in Section \ref{memorynetwork} for dephasing and depolarization to the qubits of the Bell pairs. The waiting time in the QMs from Eq.~(\ref{tB_i}) and (\ref{tC_i})
govern the dephasing noise. We define $\hat{\Phi}^{\pm}, \hat{\Psi}^{\pm}$ as the projectors onto the eigenspaces of the well known Bell pairs $\Phi^{\pm}, \Psi^{\pm}$ defined by the state vectors
\eqn{
&\ket{\Phi^+}=\frac{1}{\sqrt{2}}(\ket{0,0}+\ket{1,1}),\\
&\ket{\Phi^-}=\frac{1}{\sqrt{2}}(\ket{0,0}-\ket{1,1}),\\
&\ket{\Psi^+}=\frac{1}{\sqrt{2}}(\ket{0,1}+\ket{1,0}),\\
&\ket{\Psi^-}=\frac{1}{\sqrt{2}}(\ket{0,1}-\ket{1,0}),
\label{bellstates}}
to express the noisy resource state shared by the central station $\tilde C_i$ and the $i$-th Bob $B_i$. The noisy resource Bell pair then reads
\eqn{
\rho_{\tilde{C},B_i}&=\Big((1-f_{D})A_i+\frac{f_{D}}{4}\Big)\hat{\Phi}^+_{\tilde{C}_i,B_i}\nonumber\\
&+\Big((1-f_{D})B_i+\frac{f_{D}}{4}\Big)\hat{\Phi}^-_{\tilde{C}_i,B_i}\nonumber\\
&+\frac{f_{D}}{4}\Big(\hat{\Psi}^+_{\tilde{C}_i,B_i}+\hat{\Psi}^-_{\tilde{C}_i,B_i}\Big)
\label{noisybellpair}.}
Here, $A_i$ and $B_i=1-A_i$ are parameters taking into account the memory dephasing via
\eqn{
A_i&:=\frac{1}{2}\bigg(1+(\exp{\frac{-t_{B_i}}{T_2}})(\exp{\frac{-t_{\tilde C_i}}{T_2}})\bigg), \\\
B_i&:=\frac{1}{2}\bigg(1-(\exp{\frac{-t_{B_i}}{T_2}})(\exp{\frac{-t_{\tilde C_i}}{T_2}})\bigg).
\label{ABdefinition}}
We have not modelled imperfections during the teleportation measurement additionally as sufficiently high success probabilities have been observed in experiments (99\% in Ref.~\cite{Benhelm2008}). Alternatively a higher channel depolarization can be chosen to incorporate entanglement swapping imperfections, both of them act as depolarization channels\cite{Luong:2016}. However it has to be kept in mind that this depolarization would increase exponentially in player number. In bQSS one link is established per round, the relevant state is therefore (\ref{rhoACi}) with only one qubit $C_i$ and one single Bell pair.
In mQSS the relevant state at the hub before the entanglement swapping is again (\ref{rhoACi}), however in that case we have $N-1$ qubits, which are all in the state $C_i$. Also, instead of one single Bell pair, $N-1$ Bell pairs $\tilde{C}_1,B_1,\dots ,\tilde{C}_{N-1},B_{N-1}$ have to be established between the central station and every single Bob. The overall state in that case reads
\eqn{\rho=\rho_{A,C_1,\dots,C_{N-1}}\otimes \bigotimes_{i=1}^{N-1} \rho_{\tilde{C}_i,B_i}\label{mQSS_rhototal}.}
Before turning to the entanglement swapping, it is helpful to realise that the final state shared by Alice and the Bobs can be expressed in the GHZ basis \cite{Epping:2016hc}
\eqn{\rho_{A,B_i}=a\ket{\psi_0^+}\bra{\psi_0^+}+b\ket{\psi_0^-}\bra{\psi_0^-}\nonumber\\
+\sum_{j=1}^{2^{N-1}-1} \sum_{\sigma=+,-} c_j^{\sigma}\ket{\psi_j^{\sigma}}\bra{\psi_j^{\sigma}},
\label{GHZbasis}}
where  
\eqn{\ket{\psi_j^{\pm}}:=\frac{1}{\sqrt{2}}(\ket{0}\ket{j}\pm \ket{1}\ket{\bar{j}}),\label{GHZinbasis}}
and $j$ is a binary bit string and $\bar{j}$ its binary negation. Note here that the final state can indeed be expressed in the GHZ basis as the depolarization and dephasing are both diagonal in the GHZ-basis. Given this state, the error rates can be deduced from the prefactors in Eq.~(\ref{GHZbasis}). Every term except for $\ket{\psi_0^{\pm}}\bra{\psi_0^{\pm}}$ will deterministically yield a $Z$-error, as there is always at least one Bob whose outcome is discordant to Alice's. To find a general expression for the $X$-error, we can make use of the fast that $c_j^{+}=c_j^{-}$. The terms with $\sigma=-$ will always lead to a $X$-error, terms with $\sigma=+$ will not. The error rates can then be expressed in terms of the prefactors as
\eqn{Q^{\text{mQSS}}_{X}&=&\frac{1}{2}(1-a+b)\label{QBER_multi_X},\\
Q^{\text{mQSS}}_{A,B_i}&=&1-a-b\label{QBER_multi_Z}.}
To successfully swap the entanglement to the end nodes $B_1,\dots,B_{N-1}$, Eq.~(\ref{mQSS_rhototal}) is projected into 
\eqn{\bigotimes_{i=1}^{N-1} \bra{\phi^+}_{C_i,\tilde{C}_i}
\label{mQSS_projection}.}
The key point here is that all mixed terms in Eq.~(\ref{mQSS_projection}) will yield zero in the projection measurement as all the $C_i$'s in Eq.~(\ref{rhoACi}) will always be in the same state. Only the terms $\bra{0\dots0}_{C_1,\tilde{C}_1,\dots,C_{N-1},\tilde{C}_{N-1}}$ and $\bra{1\dots1}_{C_1,\tilde{C}_1,\dots,C_{N-1},\tilde{C}_{N-1}}$ need to be explicitly calculated. Precisely, for every term in Eq.~(\ref{rhoACi}), there are combinations in the product of Eq.~(\ref{noisybellpair}) yielding $\ket{\psi_0^{\pm}}\bra{\psi_0^{\pm}}$. The possible combinations grow in the player number $N$. However, the symmetry of the state can be used to derive analytic expressions for an arbitrary player number $N$. The key point here is to keep track of the parity of the involved terms. The needed prefactors $a$ and $b$ from Eq.~(\ref{QBER_multi_X}) and Eq.~(\ref{QBER_multi_Z}) can then be expressed as 
\eqn{a = \left( 1 - \frac{3f_{\mathrm{D}}}{4} \right) \alpha 
    + \frac{f_{\mathrm{D}}}{4} \beta 
    + 2^{N-1} \left( \frac{f_{\mathrm{D}}}{4} \right)^{N-1}}
and
\eqn{b = \left( 1 - \frac{3f_{\mathrm{D}}}{4} \right) \beta 
    + \frac{f_{\mathrm{D}}}{4} \alpha 
    + 2^{N-1} \left( \frac{f_{\mathrm{D}}}{4} \right)^{N-1}}
$\alpha$ and $\beta$ can be calculated as:
\eqn{\alpha = \mathbb{E}\left[\sum_{\substack{S \subseteq \{1, \dots, N_B\} \\ |S| \text{ even}}} \prod_{i \in S} \varphi_i \prod_{j \notin S} \vartheta_j  \right] \label{expected_dephasing_1}} ,
\eqn{\beta = \mathbb{E} \left[ \sum_{\substack{S \subseteq \{1, \dots, N_B\} \\ |S| \text{ odd}}} \prod_{i \in S} \varphi_i \prod_{j \notin S} \vartheta_j. \right]  \label{expected_dephasing_2}
}
$\vartheta>0$ and $\varphi>0$ are the coefficients in Eq.~(\ref{noisybellpair}) taking into account the memory dephasing, which is dependent on the distances $d_A$ and $d_B$. Explicitly, they read
\eqn{\vartheta_i=(1-f_{D})A_i+\frac{f_{D}}{4} ,\\ \varphi_i=(1-f_{D})B_i+\frac{f_{D}}{4}.
\label{theta_phi_def}}

\section{Composable, finite-size security proof}\label{finite}

In this section, we present the details of the security proof used to derive the results presented in the main text. The composable, finite-size results for a CKA protocol have been derived in Ref.~\cite{Grasselli:2018gk}. For secret sharing we will adapt the general results from Ref.~\cite{ourqss} to the modified GHZ protocol presented in this work. The two protocols are experimentally dual to one another in the sense that for CKA the key is derived from measurements made by Alice in the $Z$ basis whereas in QSS the key comes from measurements in the $X$ basis. They also differ in their goals. In CKA the key must be independently recoverable by each individual Bob, but all Bobs,
\eqn{\mathcal{B} = \{B_1,B_2,\ldots,B_n\}}
are assumed trustworthy so the only opponent is Eve. Thus, from a correctness perspective the relevant final strings belong to Alice and each individual Bob after all post-processing ($\mathbf{S}_A, \{ \mathbf{S}_{B_i}\}$) and from a secrecy perspective, the final state of interest at the end of the protocol is
\eqn{\rho_{\bm{S}_{A},E} = \sum_{\bm{s}_A}p(\bm{s}_A) \ket{\bm{s}_A}\bra{\bm{s}_A}\otimes \rho_{E}^{\bm{s}_A} \label{CKA},}
and we can define $\varepsilon$-security as follows.

\begin{Definition}[Conference key agreement
\cite{Grasselli:2018gk,Portmann:2022io,Renner:2005p464}]
A conference key agreement scheme as defined in Protocol~\ref{Prot} that outputs a state of the form (\ref{CKA}) is
\begin{itemize}
\item {\it $\varepsilon_c$-correct} if 
\eqn{p_{\mathrm{pass}} \mathrm{Pr}[\exists \hspace{1mm} B_i: \bm{S}_A \neq \bm{S}_{B_i}]  \leq \varepsilon_c} 
and 
\item $\varepsilon_s$-secret if 
\eqn{\hs p_{\mathrm{pass}} D\bk{\rho_{\bm{S}_{A,E}}, \tau_{\bm{S}_A}\otimes\sigma_{E}} \leq \varepsilon_s \label{sec_cka}}
where $D(\rho,\sigma) = \half ||\rho-\sigma||
_1$ is the trace distance and $\tau_{\bm{S}_A}$ is the uniform 
(i.e., maximally mixed) state over $\bm{S}_A$.
\end{itemize}
A protocol is {\it ideal} if it satisfies $\varepsilon_c = \varepsilon_s=0$ and it is called $\varepsilon_{\mathrm{sec}}$-secure if  $\varepsilon_{\mathrm{sec}}=\varepsilon_c + \varepsilon_s$. 
\label{CKAdef}
\end{Definition}
A QKD protocol is $\varepsilon$-secure if it satisfies precisely this definition but with only a single Bob \cite{Tomamichel:2012p7120,Portmann:2022io}.

By contrast, in an ($N,k$)-threshold secret sharing protocol, 
the key only needs to be reconstructed by a trusted subset of the 
set of players given by
\begin{equation}
\mathcal{T} := \{T_1,T_2,\ldots,T_{\binom{n}{k}}\}
\end{equation}
for which $T_1 := \{B_1,B_2,\ldots,B_k\}$ working collaboratively. However, unauthorised or untrusted subsets of $(k-1)$ players 
\begin{equation}
\mathcal{U} := \{U_1,U_2,\ldots,U_{\binom{n}{k-1}} \}
\end{equation}
may also be collaborating with the eavesdropper. The identity of the malicious players is completely unknown, however their number is upper bounded by $k-1$, and there are $\binom{n}{k-1}$ sets of players as explained in Section \ref{secret_sharing_rates}. Note that each untrusted subset automatically defines a complementary subset
\begin{equation}
\mathcal{C} := \{C_1,C_2,\ldots,C_{\binom{n}{k-1}} \} := \{\mathcal{B}\setminus U_1, \mathcal{B}\setminus U_2,\ldots,\mathcal{B}\setminus U_{\binom{n}{k-1}} 
\end{equation}
 
 The relevant final strings from a correctness perspective are those of Alice and any trusted subset ($\mathbf{S}_A, \{ \mathbf{S}_{T_i}\}$) and the relevant state from a secrecy perspective now includes the untrusted set in collaboration with Eve and reads
\eqn{\rho_{\bm{S}_{A},E,U_j} = \sum_{\bm{s}_A}p(\bm{s}_A) \ket{\bm{s}_A}\bra{\bm{s}_A}\otimes \rho_{E,U_j}^{\bm{s}_A} .\label{S}}
We can now define the $\varepsilon$-security of such a scheme in the composable framework.

\begin{Definition} [Secret sharing scheme \cite{ourqss}]
A secret sharing scheme as defined in Protocol~\ref{Prot} that outputs a state of the form (\ref{S}) is
\begin{itemize}
\item {\it $\varepsilon_c$-correct} if 
\eqn{\max_i \left\{ \mathrm{Pr}[\bm{S}_A \neq \bm{S}_{T_i}] \right \} \leq \varepsilon_c} 
and 
\item $\varepsilon_s$-secret if 
\eqn{\hs \max_j \left\{p_{\mathrm{pass}} D\bk{\rho_{\bm{S}_{A,E,U_j}}, \tau_{\bm{S}_A}\otimes\sigma_{E,U_j}} \right \} \leq \varepsilon_s \label{sec}}
where $D(\cdot,\cdot)$ is the trace distance and $\tau_{\bm{S}_A}$ is the uniform 
(i.e., maximally mixed) state over $\bm{S}_A$.
\end{itemize}
A protocol is {\it ideal} if it satisfied $\varepsilon_c = \varepsilon_s=0$ and it is called $\varepsilon_{\mathrm{sec}}$-secure if  $\varepsilon_{\mathrm{sec}}=\varepsilon_c + \varepsilon_s$. 
\label{SSdef}
\end{Definition}
 
 Here we will quantify the finite size of a given protocol by specifying a block size, $m$, of the number of detections in the key-generating basis which are represented by random variables $\mathbf{X}^m$ for QSS and $\mathbf{Z}^m$ for CKA. Both CKA \cite{Grasselli:2018gk} and QSS \cite{ourqss} analyses rely on entropic uncertainty relations as the key ingredient \cite{Tomamichel:2011p461}. These can be used to bound the eavesdroppers smooth conditional min-entropy $\hmin^\varepsilon(\mathbf{Y}^m|E)_{\rho_{\mathbf{Y}^mE}}$ of some random variable, $\mathbf{Y}^m$, generated by $m$ measurements of the key generating observable for either protocol. The conditional min-entropy quantifies the number of extractable bits that will appear random to an eavesdropper holding system $E$ in the sense of Eq.~(\ref{sec_cka}) and (\ref{sec}) via the 
 \emph{leftover hashing lemma} \cite{Renner:2005p464,Tomamichel:2017et}. More concretely, for a cq-state $\rho_{\mathbf{Y}_{A,E}}$ the quantum conditional min-entropy is given by
 \begin{equation}\hmin(\mathbf{Y}^m|E)_{\rho_{\mathbf{Y}_{A,E}}} = -\log_2\bk{\sup_{\{ E_\mathbf{y}\} } \sum_\mathbf{y} p(\mathbf{y})\mathrm{tr} \left \{E_\mathbf{y}\rho_E^\mathbf{y} \right \} }
 \end{equation}
 where the supremum is taken over all POVMs on the $E$ system. We can also define some other entropic quantities that will be useful later in our analysis. Firstly, we can define a dual quantity, the so-called max-entropy, by considering a state vector $\ket{Y_{A,B,E}}$ that purifies $\rho_{\mathbf{Y}_{A,E}}$ and the corresponding marginal state $\rho_{\mathbf{Y}_{A,B}} = \mathrm{tr}_E(\ket{A,B,E}\bra{A,B,E})$. The max-entropy is then defined as
\eqn{\hmax(\mathbf{Y}^m|E)_{\rho_{\mathbf{Y}_{A,E}}} = -\hmin(\mathbf{Y}^m|B)_{\rho_{\mathbf{Y}_{A,B}}}}
The \emph{smoothed} quantum conditional min- and max-entropy ($\hmin^\varepsilon(\mathbf{Y}^m|E), \hmax^\varepsilon(\mathbf{Y}^m|E)$) are given by taking the optimisations over all states that are $\varepsilon$-close to $\rho_{\mathbf{Y}_{A,E}}$ in purified distance, 
\begin{equation}
P(\rho,\sigma) := \sqrt{1- F^2(\rho,\sigma)}, 
\end{equation}
with $F(\rho,\sigma) := \mathrm{tr} \left \{| \sqrt{\rho}\sqrt{\sigma}| \right \}$ being the standard mixed-state fidelity as
\eqn{\hmin^\varepsilon(\mathbf{Y}^m|E)_{\rho_{\mathbf{Y}_{A,E}}} \ee \sup_{\tilde{\rho} \in \mathcal{P}^\varepsilon\bk{\rho_{\mathbf{Y}_{A,E}}}} \hmin(\mathbf{Y}^m|E)_{\tilde{\rho}} \nonumber ,\\
\\
\hmax^\varepsilon(\mathbf{Y}^m|E)_{\rho_{\mathbf{Y}_{A,E}}} \ee \min_{\tilde{\rho} \in \mathcal{P}^\varepsilon\bk{\rho_{\mathbf{Y}_{A,E}}}} \hmax(\mathbf{Y}^m|E)_{\tilde{\rho}} \nonumber\\ \label{smoothdef},} 
where $\mathcal{P}^\varepsilon\bk{\rho_{\mathbf{Y}_{A,E}}} = \{\rho \hspace{1mm} | P(\rho,\rho_{\mathbf{Y}_{A,E}}) \leq \varepsilon\}$. A final entropic quantity that will be necessary is the Renyi-entropy of order zero, defined for classical distributions $P_{X,Y}$ as 
\eqn{H_0(X|Y)_{P_{X,Y}} \ee \log_2 \max_{y} |\mathrm{supp}(P_X^y)|}
where $\mathrm{supp}(P_X^y)$ is the support of the conditional probability distribution of $X$ given $Y=y$. Similarly, 
one can define a smoothed entropy with respect to the purified distance as
\eqn{H_0^\varepsilon(X|Y)_{P_{X,Y}} := \min_{\tilde{Q}_{X,Y} \in \mathcal{P}\bk{P_{X,Y}}} H_0(X|Y)_{\tilde{Q}_{X,Y}} .}
Note that here, and for the rest of this work, we will suppress the subscripts on entropies for brevity.

 Returning to the security analysis, the \emph{leftover hashing lemma} then states that, after a two-universal hashing function has been applied to $\mathbf{Y}$ to obtain new strings $\mathbf{S}$ it then holds that \cite{Renner:2005p464,Tomamichel:2012p7120}
\eqn{D(\rho_{\mathbf{S}_{A,E}}, \tau_{\bm{S}_A}\otimes\sigma_{E'}) \leq 2^{-\frac{1}{2} \bk{\hmin^\varepsilon(\mathbf{Y}^m_A|E) - \ell +2}} + 2\varepsilon\nn \\ \label{hash}}
where the system $E'$ represents all eavesdropper information including additional leakage during classical communication for error correction.
Choosing $\varepsilon = \varepsilon_{\mathrm{PE}}/p_{\mathrm{pass}}$ in this expression and 
\eqn{\ell = \hmin^{\frac{\varepsilon_{\mathrm{PE}}}{p_{\mathrm{pass}}}}(\mathbf{Y}^m_A|E) + 2 +2\log_2\bk{\frac{\varepsilon_{\mathrm{PA}}}{p_{\mathrm{pass}}}} \label{lextract}}
for constants $\varepsilon_{\mathrm{PA}},\varepsilon_{\mathrm{PE}}>0$ then the right hand side of Eq.~(\ref{hash}) becomes $(\varepsilon_{\mathrm{PA}} + 2\varepsilon_{\mathrm{PE}})/p_{\mathrm{pass}}$. The next step is to realise that the total eavesdropper system can be divided into two parts $E' = E,R$, where $R$ is a register containing all information leaked during error correction and $E$ is a quantum system that purifies the relevant shared information between the legitimate parties. If we define the leaked information for a $\varepsilon_c$-correct EC protocol as  the $\ell_{\mathrm{EC}}$ such that 
\begin{equation}
\hmin^{\varepsilon}(\mathbf{Y}^m_A|ER) = \hmin^\varepsilon(\mathbf{Y}^m_A|E) -\ell_{\mathrm{EC}}, 
\end{equation}
then, recalling that $p_{\mathrm{pass}}<1$, we can combine Eq.~(\ref{lextract}) with Eq.~(\ref{smoothdef}) to obtain a lower bound for the extractable key length that is $(\varepsilon_{\mathrm{PA}} + 2\varepsilon_{\mathrm{PE}})$-secret and $\varepsilon_c$-correct of
\eqn{\ell = \hmin^{\frac{\varepsilon_{\mathrm{PE}}}{p_{\mathrm{pass}}}}(\mathbf{Y}^m_A|E) + 2 - \ell_{\mathrm{EC}} -2\log_2\bk{\frac{1}{\varepsilon_{\mathrm{PA}}}} \label{lextract2}.}

The goal of the security analysis is then to estimate the conditional min-entropy and information leakage appropriate for the CKA and QSS protocols.
Considering first the information leakage during EC, there are two processes by which information can be leaked - either through the EC coding itself or the hashing based check used to certify the $\varepsilon_c$-correctness of the outputs. The parameter $\varepsilon_c$ is independent of the code choice in the sense that, for whatever EC code is chosen, if a hash test is passed we can be sure the final string is $\varepsilon_c$-correct. 

If a code is chosen that is insufficient for the channel noise, this causes the check to fail with high probability. This concept is captured by the concept of $\varepsilon_{\mathrm{rob}}$-robustness of a given EC code, which means it will pass the hash tests with probability $1-\varepsilon_{\mathrm{rob}}$. In an experiment, 
the total leakage can be quantified by simply counting the number of transmitted bits for whichever EC protocol and check are used, but we can also bound the performance of an optimal EC protocol using the following result,

\begin{Theorem}[Adapted from Ref.~\cite{Grasselli:2018gk}]
    Given a probability distribution $P_{\mathbf{Y}_A\mathbf{Y}_{\mathcal{B}}}$ between Alice and a set of $N$ Bobs, $\mathcal{B} = \{B_1,
    \dots ,B_i,
    \dots,B_N \}$, there exists a one-way EC protocol, that is,
    $\varepsilon_{\mathrm{EC}}$-correct,
$2(N - 1)\varepsilon'$-robust on $P_{\mathbf{Y}_A\mathbf{Z}_{\mathcal{B}}}$, and has leakage
\eqn{\ell_{\mathrm{EC}} \leq \max_iH_0^{\varepsilon'}(\mathbf{Y}_A|B_i) + \log_2\bk{\frac{2(N-1)}{\varepsilon_{\mathrm{EC}
}}}. \label{lEC}}
\end{Theorem}

Turning to the information gained by the eavesdropper through the noisy channel, we can exploit an entropic uncertainty relation for the smoothed conditional min- and max-entropies. Given a tripartite state vector $\ket{A,B ,E}$ which can be assumed pure without loss of generality and the possibility to measure in either the Pauli $X$ or $Z$ basis on the $A$ system, and the key relation for an $m$-round protocol reads \cite{Tomamichel:2011p461},
 \eqn{\hmin^\varepsilon(\mathbf{X}^m_A|E) + \hmax^\varepsilon(\mathbf{Z}^m_A|B) \geq m.
 \label{EUR}}
Another tool we will need to evaluate these bounds is a method of bounding Renyi entropies 
via Shannon entropies which can be achieved as follows.

\begin{Lemma}[Adapted from Refs.~\cite{Tomamichel:2012p7120,Grasselli:2018gk}]
    Let $P_{\mathbf{Y}_A^n\mathbf{Y}_B^n}$ be a probability distribution on classical variables describing correlated $n$-bit strings. If the number of discordant (i.e., non-matching) bits, $Q_{A,B}^n$, is probabilistically bounded by
    \eqn{\mathrm{Pr}\left [Q_{A,B}^n > Q_{A,B}+\xi \right] \leq \varepsilon^2 \label{statcond}}
    then it holds that \eqn{\hmax^\varepsilon(\mathbf{Y}_A^n|\mathbf{Y}_B^n) \leq n h_2(Q_{A,B}+\xi ) \label{hmaxbnd}}
    and  \eqn{H_0^\varepsilon(\mathbf{Y}_A^n|\mathbf{Y}_B^n) \leq n h_2(Q_{A,B}+\xi). \label{h0bnd}}
\end{Lemma}

Finally, we need two kinds of statistical bounds. The first kind are Hoeffding bounds, which can be applied when the sample mean of a distribution is known and we wish to bound the probability of a certain number of events occurring in a number of samples. For example, the number of observed, $Q_Y^m$ errors after $m$ transmissions through a qubit channel with a true QBER, $Q_{A,B}$, will satisfy
    \eqn{\mathrm{Pr}\left [Q_{A,B}^m > Q_{A,B}+\xi_1(m,\varepsilon) \right] &\leq& \varepsilon^2 \nn \\
    \mathrm{with} \hs \xi_1(\varepsilon,m) &:=& \sqrt{\frac{\log(1/\varepsilon)}{m}} .\label{xi1}}
The second kind concerns a situation where the sample mean for $n$ rounds is unknown, but is estimated by randomly choosing $k$ samples for a total of $N = m+k$ rounds. This sampling without replacement situation can be analysed via Serfling's bound and can be shown that if an error ratio of $Q^k$ is observed on the $k$ samples then we can probabilistically bound the error that would be observed on the remaining $m$ samples as \cite{Serfling:1974dx,Tomamichel:2012p7120},
    \eqn{\mathrm{Pr}\left [Q_{Y}^m > Q_{Y}^k + \xi_2(\varepsilon,m,k) \right] 
    \leq \varepsilon^2 }
    with
    \eqn{\hs \xi_2(\varepsilon,m,k) := \sqrt{\frac{(m+k)(k+1)}{mk^2}\ln(1/\varepsilon)}. \label{stat2} }
We can now put all of this together to establish the expected amount of secure, extractable key for both CKA and QSS protocols.

\begin{widetext}
\begin{Theorem}[Conference key agreement]
An ($N,L,r,\varepsilon_c,\varepsilon_s,\varepsilon_{\mathrm{rob}},m,k,\{\tilde{Q}_{Z_A|B_i}\},Q_{X_A|\mathcal{B}}$)-conference key agreement protocol as defined in Protocol~\ref{Prot} and carried out via L uses of a network yielding $m$ Z-basis key generation detections, $k$ $X$-basis check detections, a $Z$-basis QBER between Alice and each $B_i \in \mathcal{B}$ pre-characterised to be $\{\tilde{Q}_{Z_A|B_i}\}$ and a measured $X$-basis QBER $Q_{X_A|\mathcal{B}}$ is $\varepsilon_{\mathrm{rob}}$-robust, $(2\varepsilon_{\mathrm{PE}} + \varepsilon_{\mathrm{PA}})$-secret and $\varepsilon_c$-correct with an extractable key length of 
 \eqn{\ell_{\mathrm{CKA}} &\geq& m - m h_2(Q_{X_A|\mathcal{B}} + \xi_2(\varepsilon_{\mathrm{PE}},m,k))  
    - \max_iH_0^{\varepsilon_{\mathrm{rob}}}(\mathbf{Z}^m_A|B_i) - \log_2\bk{\frac{(N-1)}{2\varepsilon_{\mathrm{c}} \varepsilon^2_{\mathrm{PA}}}} 
    - L \cdot h_2(r) ,\label{lcka_thm} }
conditioned on the protocol not aborting. For an honest implementation where the $Z$-basis QBER is the same as the pre-characterised value the expected performance is given by
\eqn{\EV{\ell_{\mathrm{CKA}}} \ee  (1-\varepsilon_{\mathrm{rob}})\left [\EV{m}\bk{1-h_2\bk{\EV{Q_{X_A|X_{\mathcal{B}}}} + \xi_2(\varepsilon_{\mathrm{PE}},\EV{m},\EV{k})} - \max_i h_2\bk{\tilde{Q}_{Z_A|{B_i}} + \xi_1(\varepsilon_\mathrm{z},\EV{m})} } -L \cdot h_2(r) \right. \nn  \\
 &-& \left. \log_2\bk{\frac{(N-1)}{2\varepsilon_{\mathrm{c}} \varepsilon_{\mathrm{PA}}^2}} \right ] \label{lCKAEV}}
where
$\varepsilon_z := 
\varepsilon_{\mathrm{rob}}/\sqrt{N-1}$.
\end{Theorem}

\emph{Proof:} Let $m$ ($k)$ be the number of Alice's $Z$-basis ($X$-basis) measurements for key generation (parameter estimation) and $\mathbf{Z}^m_A$ ($\mathbf{X}^k_A$) the corresponding random variable. The amount of $\varepsilon_{\mathrm{PE}} + 2 \varepsilon_{\mathrm{PA}}$-secret key can be calculated by applying 
Eq.\ (\ref{lextract2}) to the output of the CKA protocol conditioned on all tests passing which yields
\eqn{\ell_{\mathrm{CKA}} \ee \hmin^{\frac{\varepsilon_{\mathrm{PE}}}{p_{\mathrm{pass}}}}(\mathbf{Z}^m_A|E) - \ell_{\mathrm{EC}} - L \cdot h_2(r) 
-2\log_2\bk{\frac{1}{\varepsilon_{\mathrm{PA}}}} +2,}
where the $L \cdot h_2(r) $ term accounts for the fact that the CKA protocol uses a pre-shared key which must be replenished from the generated key.
We then apply the entropic uncertainty relation Eq.~(\ref{EUR}) to the tripartite system made up of Alice, Eve and the set of all Bobs, $A,\mathcal{B},E$, conditioned upon all tests having been passed, which gives
\eqn{\ell_{\mathrm{CKA}} \geq m - \hmax^{\frac{\varepsilon_{\mathrm{PE}}}{p_{\mathrm{pass}}}}(\mathbf{X}^m_A|\mathcal{B}) - \ell_{\mathrm{EC}} - L \cdot h_2(r)  
-2\log_2\bk{\frac{1}{\varepsilon_{\mathrm{PA}}}} +2 .\label{cka1}}
Following Ref.~\cite{Tomamichel:2012p7120}, we note that, based upon the observed $X$-basis statistics, Bayes theorem says that the counterfactual probability of errors that would have been observed had $X$-measurements been made in the $Z$-rounds is bounded by
    \eqn{ \mathrm{Pr}\left [Q_{X}^m > Q_{X}^k + \xi_2(\varepsilon_{\mathrm{PE}},m,k)|\mathrm{pass} \right]  
    = \frac{\mathrm{Pr}\left [Q_{X}^m > Q_{X}^k + \xi_2(\varepsilon_{\mathrm{PE}},m,k) \right]}{p_{\mathrm{pass}}} \leq \frac{\varepsilon_{\mathrm{PE}}^2}{p_{\mathrm{pass}}} \leq  \frac{\varepsilon_{\mathrm{PE}}^2}{p_{\mathrm{pass}}^2} }
where we have applied Eq.~(\ref{stat2}) in the first inequality and used that $p_{\mathrm{pass}}<1$ in the second. Comparing this with Eq.~(\ref{statcond}) we can now apply Eq.~(\ref{hmaxbnd}) to bound the max-entropy term in Eq.~(\ref{cka1}) to obtain
    \eqn{\ell_{\mathrm{CKA}} &\geq& m - m h_2(Q_{X_A|\mathcal{B}} + \xi_2(\varepsilon_{\mathrm{PE}},m,k))  
    - \ell_{\mathrm{EC}} - L \cdot h_2(r) -2\log_2\bk{\frac{1}{\varepsilon_{\mathrm{PA}}}} +2 . \label{cka2}}
We can then apply Eq.~(\ref{lEC}) which bounds the information leaked by $\varepsilon_c$-correct and $\varepsilon_\mathrm{rob}$-robust EC code. Note that this $\varepsilon_\mathrm{rob}$ parameter does not effect the overall correctness or secrecy of the protocol (which is conditioned on the EC checks passing) but only effects the overall rate at which EC checks will pass on average. In other words, a poor choice of code which has been  insufficient to correct the errors in an implementation might result in a high probability of aborting but would not lead to a security breach. For this reason, in most QKD implementations the QBER in the key generation need not be measured in real-time, but can be estimated offline and used to choose an EC code without compromising security. This is how we proceed in our protocol, although it is also possible to carry out real-time estimation at the price of sacrificing further data for parameter estimation (e.g., 
this approach is taken in Ref.~\cite{Grasselli:2018gk}). Substituting in Eq.~(\ref{lEC}) and collecting logarithmic terms we find an $\varepsilon_{\mathrm{PE}} + 2 \varepsilon_{\mathrm{PA}}$-secret, $\varepsilon_c$-correct and $\varepsilon_{\mathrm{rob}}$-robust key can be extracted of length
    \begin{equation}
    \ell_{\mathrm{CKA}} \geq m - m h_2(Q_{X_A|\mathcal{B}} + \xi_2(\varepsilon_{\mathrm{PE}},m,k))  
    - \max_iH_0^{\varepsilon_{\mathrm{rob}}}(\mathbf{Z}^m_A|B_i) - \log_2\bk{\frac{(N-1)}{2\varepsilon_{\mathrm{c}} \varepsilon^2_{\mathrm{PA}}}} 
    - L \cdot h_2(r) . \label{cka3}
    \end{equation}
We are now interested in calculating the expected value of this length for an honest implementation where the average $Z$-QBER during key generation for any $B_i$ is in fact same as the quantity $Q^i_{Z_A|Z_{B}}$ estimated previously over the network. Again, recall that in a real implementation no assumption is made regarding the true $Z$-basis QBER -- if Eve were to dynamically increase it there would be no security leak, the EC protocol would simply abort with high probability. Here we will calculate the expected performance. In order for all EC checks to pass we must consider the passing probability for all Bob's simultaneously (alternatively, the probability that any Bob observes an unexpectedly high QBER). The probability of any single Bob obtaining a QBER for the $m$ key generating rounds that is larger than $\tilde{Q}_{Z_A|B_i} + \xi_1(m,\varepsilon_{\mathrm{rob}})$ can be obtained from Eq.~(\ref{xi1}). Applying this bound to each Bob we immediately have
\eqn{\mathrm{Pr}\left [Q^m_{Z_A|B_i} > \max_i \tilde{Q}_{Z_A|B_i}+\xi_1(m,\varepsilon_z) \right] \leq \varepsilon_z^2,  \hspace{1mm} \forall i .\label{QABi}}
Turning to the $N$-party error correction problem, we need to bound the probability that any of the Bob's fail to correct their output, in other words the union of the EC failure's for each Bob. We have that
\eqn{\mathrm{Pr}\left [\bigcup_{i=1}^{N-1}  Q_{Z_A|B_i}^m > \max_i \tilde{Q}_{Z_A|B_i}+\xi_1(m,\varepsilon_z)  \right] \leq \sum_{i=1}^{N-1} \mathrm{Pr}\left [ Q_{Z_A|B_i}^m > \max_i \tilde{Q}_{Z_A|B_i}+\xi_1(m,\varepsilon_z) \right] \leq (N-1)\varepsilon_z^2,}
where we have used the union bound in the first inequality and Eq.~(\ref{QABi}) in the second. Combining this bound with Eq.~(\ref{h0bnd}) and substituting in Eq.~(\ref{cka3}), we obtain
    \eqn{\ell_{\mathrm{CKA}} &\geq& m - m h_2(Q_{X_A|\mathcal{B}} + \xi_2(\varepsilon_{\mathrm{PE}},m,k))  
    - mh_2\bk{\max_i \tilde{Q}_{Z_A|B_i}+\xi_1(m,\varepsilon_z) } - L \cdot h_2(r) 
    - \log_2\bk{\frac{(N-1)}{2\varepsilon_{\mathrm{c}} \varepsilon^2_{\mathrm{PA}}}}  \label{cka4} }
where 
\begin{equation}
\varepsilon_z := \varepsilon_{\mathrm{rob}}/\sqrt{N-1}. 
\end{equation}
The expected value of the secret fraction for a given network is simply given by $\EV{\ell_{\mathrm{CKA}}} = (1-\varepsilon_{\mathrm{rob}}) \ell_{\mathrm{CKA}}$ where $\ell_{\mathrm{CKA}}$ is evaluated using Eq.~(\ref{cka4}) and the expected values for the yields and QBER calculated previously which completes the proof. \ensuremath{\Box}

\begin{Theorem} (Secret sharing) 
An ($N,L,r,\varepsilon_c,\varepsilon_s,\varepsilon_{\mathrm{rob}},m,\{k_i\},\{Q_{Z_A|B_i}\},\tilde{Q}_{X\mathcal{B}}$)-secret sharing protocol as defined in Protocol~\ref{Prot} and carried out via $L$ uses of a network yielding $m$ $X$-basis key generation detections, $k^i$ $Z$-basis check detections with each $B_i \in \mathcal{B}$, a measured $Z$-basis QBER of $Q^{k_i}_{Z_A|B_i}$ and a pre-characterised $X$-basis QBER $\tilde{Q}_{X_A|\mathcal{B}}$ is $\varepsilon_{\mathrm{rob}}$-robust, $(2\varepsilon_{\mathrm{PE}} + \varepsilon_{\mathrm{PA}})$-secret and $\varepsilon_c$-correct with an extractable key length of 
\eqn{\ell_{\mathrm{SS}} &\geq& m - m h_2\bk{\max_i Q_{Z_A|B_i}^{k_i} + \xi_2(m,k_i,\varepsilon_z)}- \ell_{\mathrm{EC}}  
- \log_2\bk{\frac{1}{4\varepsilon_{\mathrm{PA}}^2}}.}
We then bound the EC leakage via Eq.~(\ref{lEC}), to obtain
\eqn{\ell_{\mathrm{SS}} &\geq& m - m h_2\bk{\max_i Q_{Z_A|B_i}^{k_i} + \xi_2(m,k_i,\varepsilon_z)} 
- H^{\varepsilon_{\mathrm{rob}}}_0(X_A|\mathcal{B}) - \log_2\bk{\frac{2(N-1)}{\varepsilon_{\mathrm{EC}}}}  
- \log_2\bk{\frac{1}{4\varepsilon_{\mathrm{PA}}^2}}.}
For an honest implementation where the $X$-basis QBER is stable the expected value for the extractable key length is given by
   \eqn{\EV{\ell_{\mathrm{SS}}} &\geq& (1-\varepsilon_{\mathrm{rob}})\left [\EV{m}\left [1 - h_2\bk{\max_i Q_{Z_A|B_i}^{\EV{k_i}} + \xi_2(\EV{m},\EV{k_i},\varepsilon_z)}- h_2\bk{\tilde{Q}_{X_A|\mathcal{B}} + \xi_1(\EV{m},\varepsilon_{\mathrm{rob}})}\right ] \right . \nn \\
   &-& \left. \log_2\bk{\frac{(N-1)}{2\varepsilon_{\mathrm{EC}}\varepsilon_{\mathrm{PA}}^2}}\right ].
   \label{lSSEV}}
\end{Theorem}
\emph{Proof:} Let $m$ ($k)$ be the number of Alice's $X$-basis ($Z$-basis) measurements for key generation (parameter estimation) and $\mathbf{X}^m_A$ ($\mathbf{Z}^k_A$) the corresponding random variable. Cryptographically speaking, the situation is very different from CKA. We must consider the possibility that any of the untrusted subsets (for an ($N-1,N-1$)-threshold scheme this is any $N-2$ subset of Bobs) may be collaborating with Eve and take the worst case. The amount of $\varepsilon_{\mathrm{PE}} + 2 \varepsilon_{\mathrm{PA}}$-secret key relative to any fixed malicious Bob can be calculated directly by applying Eq.\ (\ref{lextract2}) to the output of the QSS protocol conditioned on all tests passing, so the worst case extractable key is given by \cite{ourqss}
\eqn{\ell_{\mathrm{SS}} \ee \min_j\hmin^{\frac{\varepsilon_{\mathrm{PE}}}{p_{\mathrm{pass}}}}(\mathbf{X}^m_A|E,U_j) - \ell_{\mathrm{EC}} - \log_2\bk{\frac{1}{4\varepsilon_{\mathrm{PA}}^2}}.}
For each untrusted subset, we can apply the entropic uncertainty relation Eq.~(\ref{EUR}) to the tripartite system $\ket{AB_jE,U_j}$ made up of Alice, the joint system of Eve and $U_j$ and the corresponding complementary which in this case will be a single Bob which we will denote $B_j$, conditioned upon all tests having been passed which 
yields
\eqn{\ell_{\mathrm{SS}} &\geq& m - \max_j\hmax^{\frac{\varepsilon_{\mathrm{PE}}}{p_{\mathrm{pass}}}}(\mathbf{Z}^m_A|B_j) - \ell_{\mathrm{EC}} 
- \log_2\bk{\frac{1}{4\varepsilon_{\mathrm{PA}}^2}}.}
To upper bound the max-entropy term using Lemma 1 we need to bound the probability that a certain Z-basis QBER would have been observed in the $m$ key generation for any one of the Bobs, given that another QBER has  actually been observed. First, for all $i$ we have from 
Eq.~(\ref{stat2}) that
\eqn{\mathrm{Pr} 
\left [Q_{Z_A|B_i}^m >\max_i 
Q_{Z_A|B_i}^k + \xi_2(m,k_i,\varepsilon_z) \right] 
\leq \varepsilon_z^2 \label{Qzxi2}.}
Then we can immediately write
\eqn{\mathrm{Pr}\left [\bigcup_{i=1}^{N-1}  Q_{Z_A|B_i}^m > Q^k_{Z_A|B_i} + \xi_2(m,k,\varepsilon_z)| \mathrm{pass} \right] \ee \frac{\mathrm{Pr}\left [\bigcup_{i=1}^{N-1}  Q_{Z_A|B_i}^m > Q^k_{Z_A|B_i} + \xi_2(m,k,\varepsilon_z) \right]}{p_{\mathrm{pass}}} \nn \\
&\leq& \frac{\sum_{i=1}^{N-1} \mathrm{Pr} \left [Q_{Z_A|B_i}^m > Q_{Z_A|B_i}^k + \xi_2(m,k,\varepsilon_z) \right] }{p_{\mathrm{pass}}} \nn \\
&\leq& \frac{(N-1) \varepsilon_z^2}{p^2_{\mathrm{pass}}}, }
where we have used the union bound in the first inequality and Eq.~(\ref{Qzxi2}) and $p_{\mathrm{pass}}<1$ in the second. Setting $\varepsilon_z := \varepsilon_{\mathrm{PE}}/\sqrt{N-1}$ and substituting into Eq.~(\ref{statcond}), we can apply Eq.~(\ref{hmaxbnd}), to get
\eqn{\ell_{\mathrm{SS}} &\geq& m - m h_2\bk{\max_i Q_{Z_A|B_i}^{k_i} + \xi_2(m,k_i,\varepsilon_z)}- \ell_{\mathrm{EC}} 
- \log_2\bk{\frac{1}{4\varepsilon_{\mathrm{PA}}^2}}.}
We then bound leakage for an $\varepsilon_{\mathrm{rob}}$-robust and $\varepsilon_{\mathrm{EC}}$-correct EC scheme via Eq.~(\ref{lEC}), to obtain
\eqn{\ell_{\mathrm{SS}} &\geq& m - m h_2\bk{\max_i Q_{Z_A|B_i}^{k_i} + \xi_2(m,k_i,\varepsilon_z)} 
- H^{\varepsilon_{\mathrm{rob}}}_0(X_A|\mathcal{B}) - \log_2\bk{\frac{2(N-1)}{\varepsilon_{\mathrm{EC}}}}  
- \log_2\bk{\frac{1}{4\varepsilon_{\mathrm{PA}}^2}}.}
To calculate the expected value in an honest implementation where the QBER between all of the Bobs and Alice in the $X$-basis we apply Eq.~(\ref{xi1}), which yields
\eqn{\mathrm{Pr}\left [Q_{X_A|\mathcal{B}}^m \geq \tilde{Q}_{X_A|\mathcal{B}} + \xi_1(m,\varepsilon_{\mathrm{rob}}) \right] \leq \varepsilon_{\mathrm{rob}}^2.}
Combining this with Eq.~(\ref{h0bnd}), 
we finally arrive at
   \eqn{\EV{\ell_{\mathrm{SS}}} &\geq& (1-\varepsilon_{\mathrm{rob}})\left [\EV{m}\left [1 - h_2\bk{\max_i Q_{Z_A|B_i}^{\EV{k_i}} + \xi_2(\EV{m},\EV{k_i},\varepsilon_z)}- h_2\bk{\tilde{Q}_{X_A|\mathcal{B}} + \xi_1(\EV{m},\varepsilon_{\mathrm{rob}})}\right ] \right .\nn\\
   &-& \left. \log_2\bk{\frac{(N-1)}{2\varepsilon_{\mathrm{EC}}\varepsilon_{\mathrm{PA}}^2}} \right] }
which completes the proof. \ensuremath{\Box}
\end{widetext}

\section{Computing secret key lengths and additional results}\label{App:extra_results}

For the multi-partite protocols, we use the expected values of the key generation and parameter estimation basis yields and QBERs which we recapitulate here. For the CKA protocols the expected values for the $m$ $Z$-basis key generation measurements and $k$ $X$-basis, parameter estimation measurements over the bottleneck network characterised by transmissions $p_A$ and $p_B$ are
\eqn{\EV{m} \ee \eta_{\mathrm{k}}^{\mathrm{CKA}}Y_{\mathrm{m}}L = p_{\mathrm{key}}p_A L ,\\
\EV{k} \ee \eta_{\mathrm{c}}^{\mathrm{CKA}}Y_{\mathrm{m}}L =(1-p_{\mathrm{key}})p_A L,}
where we have used Eqs.~(\ref{yields_asymII}) and (\ref{eta_key_check_CKA}). For the secret sharing protocols they are given by Eq.~(\ref{eta_key_check_SS}) as
\eqn{\EV{m} \ee \eta_{\mathrm{k}}^{\mathrm{QSS}}Y_{\mathrm{m}}L = p_{\mathrm{key}}^N p_A L ,\\
\EV{k} \ee \eta_{\mathrm{c}}^{\mathrm{QSS}}Y_{\mathrm{m}}L = (1-p_{\mathrm{key}})(1-p_{\mathrm{key}}^{N-2})p_A L}
where we recall that the difference in the expected values of both schemes arises from the possibility of using a pre-shared key for CKA. The QBER for the $Z$-basis is calculated in Eq.~(\ref{QBER_multi_Z}) and for the $X$-basis in Eq.~(\ref{QBER_multi_X}). Finally, the security `budget' for a given $\varepsilon$-secure protocol has been chosen to be
\eqn{\varepsilon_{\mathrm{c}} = \frac{\varepsilon}{2}, \hs \varepsilon_{\mathrm{PA}} = \frac{\varepsilon}{4}, \hs \varepsilon_{\mathrm{PE}} = \frac{\varepsilon}{8}}
such that $\varepsilon_c + \varepsilon_{\mathrm{PA}}+2\varepsilon_{\mathrm{PE}} = \varepsilon$.

\begin{figure}[htb]
\includegraphics[width=0.46\textwidth]{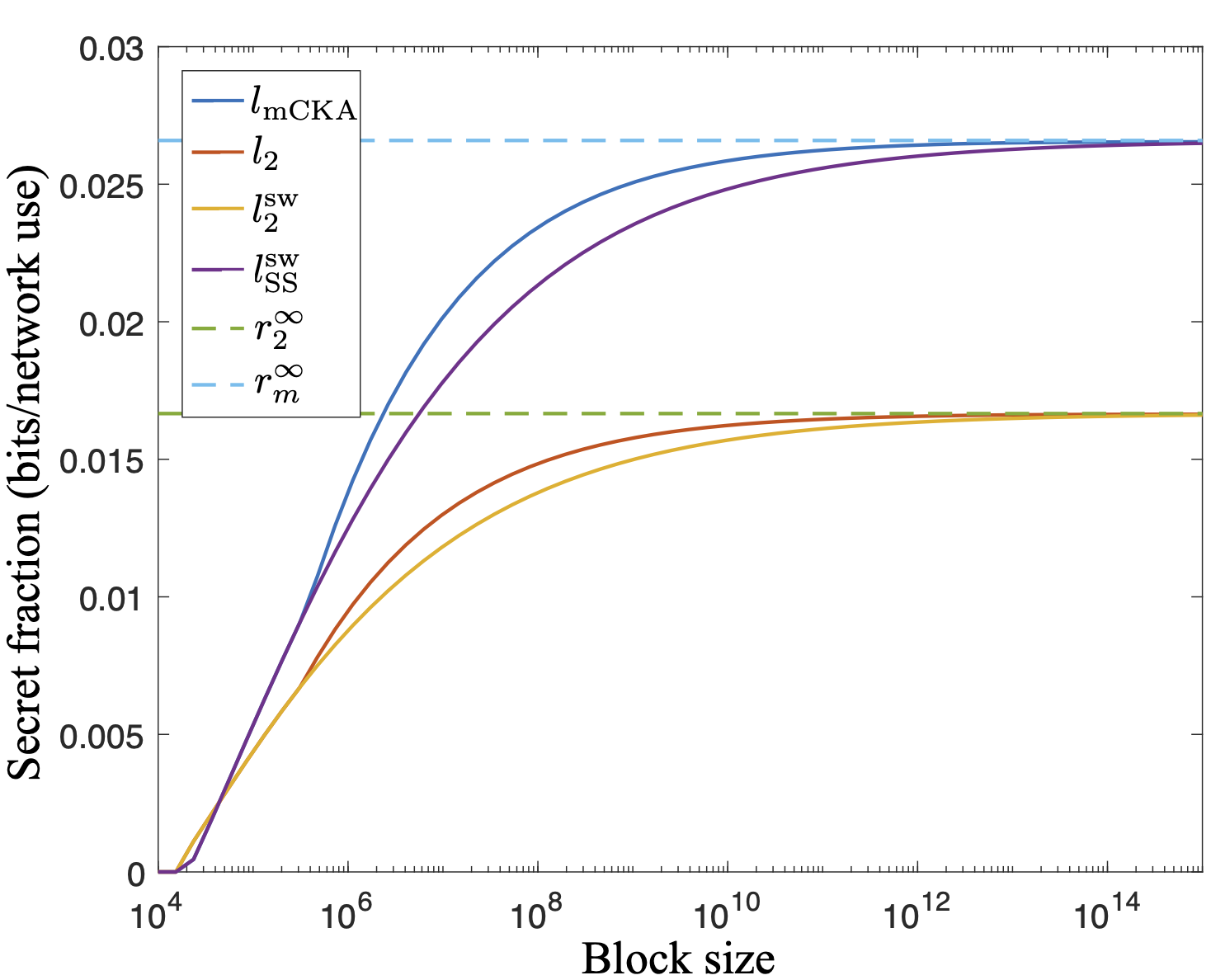}
\caption{Secret fraction as a function of block size for multi-partite entanglement based QSS protocol (solid purple) where active basis switching is essential, CKA (solid blue) and bi-partite-QKD based rate protocol for either QSS or CKA using either preshared key (solid red) or randomised basis-switching (solid yellow). Performance in the asymptotic limit for multi-partite (dashed light blue) and bi-partite protocols (dashed green) are plotted for comparison. Network parameters are $d_A = 50$ km, $d_B = 4$ km, $f_D = 0.01$ and the security parameter is $\varepsilon = 10^{-10}$. The basis probability is numerically optimised at each point.\label{fig:symm_blocksize_full}}
\end{figure}

To make a fair comparison with the bi-partite protocols, we must recall that $N-1$ QKD protocols must be performed to carry out a single round of multi-partite QSS or CKA, therefore the rates are calculated using the multi-partite formulae but with $N=2$ and the security parameter scaled such that $\varepsilon_{\mathrm{QKD}} = \varepsilon/(N-1)$. The relevant QBERs emerge from Eq.~(\ref{QBER_multi_X}) and Eq.~(\ref{QBER_multi_Z}) with $N=2$. We take the maximum over a strategy that allows for pre-shared key, so that the yields are given by
\begin{eqnarray}\EV{m} &=& \eta_{\mathrm{k}}^{\mathrm{CKA}}Y_{\mathrm{b}}L = \frac{p_{\mathrm{key}} p_A L}{N-1} ,\\
\EV{k} &=&  \eta_{\mathrm{c}}^{\mathrm{CKA}}Y_{\mathrm{b}}L = (1-p_{\mathrm{key}}) p_A L,
\end{eqnarray}
and the secure key is calculated via Eq.~(\ref{lCKAEV}), or schemes in which basis switching is instead used, so that the yields are given by
\eqn{\EV{m} \ee \eta_{\mathrm{k}}^{\mathrm{QSS}}Y_{\mathrm{b}}L = \frac{p_{\mathrm{key}}^N p_A L}{N-1}  ,\\
\EV{k} \ee \eta_{\mathrm{c}}^{\mathrm{QSS}}Y_{\mathrm{b}}L = (1-p_{\mathrm{key}})^2 p_A L,}
and the key is calculated according to Eq.~(\ref{lSSEV}). In Fig.~\ref{fig:symm_blocksize_full} we plot the secret fractions for CKA and QSS protocols along with a QKD based version with and without pre-shared key (the bi-partite curve in Fig.~\ref{fig:symm_blocksize} is given by taking the maximum of the bi-partite curve in this graph).

\begin{figure}[htb]
\includegraphics[width=0.45\textwidth]{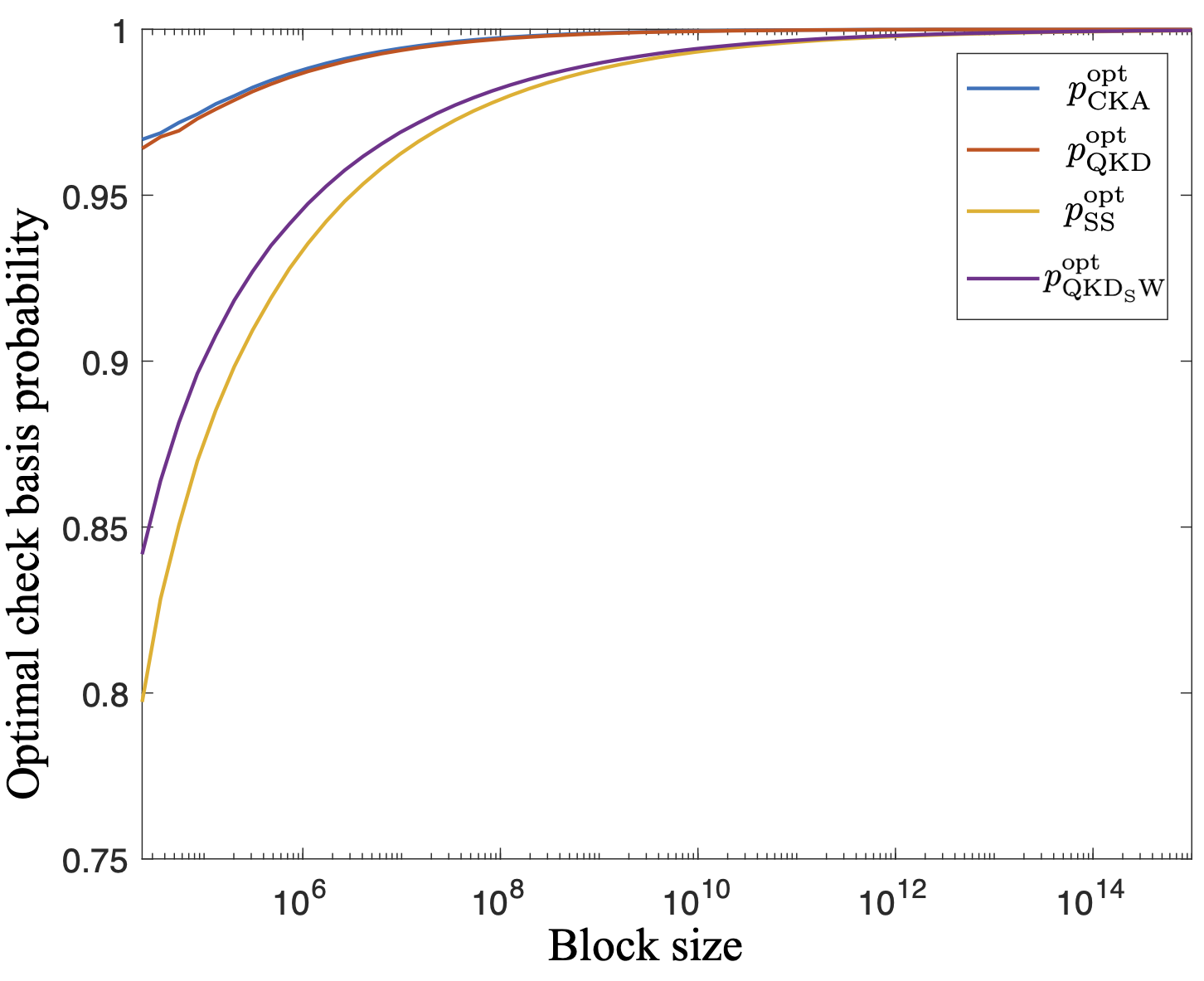}
\caption{Optimal key basis probability as a function of block size for CKA protocol using pre-shared key in the multi-partite (blue) and bi-partite (red) implementations as well as a multi-partite (yellow) and biparite (purple) QSS protocol where pre-shared key may not be used. Parameters are idential to Fig.~\ref{fig:symm_blocksize_full}.\label{fig:popt}}
\end{figure}

For all protocols, the probability for measuring the key basis is numerically optimised. The results are shown in Fig.~\ref{fig:popt}, where we see that there is a significant difference in the optimal value for the multi-partite or bi-partite strategies which is larger for small block sizes. For both the multi-partite and bi-partite case there is a small difference depending upon whether pre-shared key is used.

\vspace*{2cm}

\newpage{}
\bibliography{QSS_paper.bib}

\end{document}